\documentclass[11pt]{article}
\usepackage{amsmath,amssymb,amsfonts}
\usepackage[nosort]{cite}
\usepackage{graphicx,color}
\usepackage[colorlinks=true, linkcolor=blue, citecolor=blue, linktoc=all]{hyperref}
\usepackage[usenames,dvipsnames]{xcolor}
\usepackage{sfmath}
\usepackage{tikz-cd}
\usepackage{pict2e}

\makeatletter
\DeclareRobustCommand{\loplus}{\mathbin{\mathpalette\dog@lsemi{+}}}
\DeclareRobustCommand{\lotimes}{\mathbin{\mathpalette\dog@lsemi{\times}}}
\DeclareRobustCommand{\roplus}{\mathbin{\mathpalette\dog@rsemi{+}}}
\DeclareRobustCommand{\rotimes}{\mathbin{\mathpalette\dog@rsemi{\times}}}

\newcommand{\dog@rsemi}[2]{\dog@semi{#1}{#2}{-90,90}}
\newcommand{\dog@lsemi}[2]{\dog@semi{#1}{#2}{270,90}}
\newcommand{\dog@semi}[3]{%
  \begingroup
  \sbox\z@{$\m@th#1#2$}%
  \setlength{\unitlength}{\dimexpr\ht\z@+\dp\z@\relax}%
  \makebox[\wd\z@]{\raisebox{-\dp\z@}{%
    \begin{picture}(1,1)
    \linethickness{\variable@rule{#1}}
    \roundcap
    \put(0.5,0.5){\makebox(0,0){\raisebox{\dp\z@}{$\m@th#1#2$}}}
    \put(0.5,0.5){\arc[#3]{0.5}}
    \end{picture}%
  }}%
  \endgroup
}
\newcommand{\variable@rule}[1]{%
  \fontdimen8  
  \ifx#1\displaystyle\textfont3\else
    \ifx#1\textstyle\textfont3\else
      \ifx#1\scriptstyle\scriptfont3\else
        \scriptscriptfont3\relax
  \fi\fi\fi
}
\usepackage[makeroom]{cancel}
\usepackage{slashed,youngtab}

\newcommand{\mX}{\mathfrak{X}}

\newcommand{\teta}{{\tilde{\eta}}}

\newcommand{\un}[1]{\underline{#1}}

\newcommand{\LCnabla}{\mathring{\nabla}}

\newcommand{\chkM}{{\color{red} \,\checkmark\kern-5pt{}_{M}}}

\newcommand{\be}{\begin{equation}}
\newcommand{\ee}{\end{equation}}
\newcommand{\beq}{\begin{eqnarray}}
\newcommand{\eeq}{\end{eqnarray}}
\newcommand{\bea}{\begin{eqnarray}}
\newcommand{\eea}{\end{eqnarray}}
\newcommand{\beqn}{\begin{eqnarray}}
\newcommand{\eeqn}{\end{eqnarray}}

\def\pa{\partial}

\newcommand{\hlt}[1]{{\color{WildStrawberry}{\em #1}}\index{#1}}
\newcommand{\dM}{\mathfrak{diff}(M)}
\newcommand{\dS}{\mathfrak{diff}(S)}

\newcommand{\colr}[1]{{\color{black}#1}} 
\newcommand{\colg}[1]{{\color{black}#1}} 
\newcommand{\colb}[1]{{\color{black}#1}} 

\newcommand{\RR}{\mathbb{R}}

\newcommand{\thistitle}{Isolated Surfaces and Symmetries of Gravity}
\newcommand{\ulb}[1]{
	\centerline{
		\begin{minipage}[c]{0.7\textwidth}
			\begin{center}
			${}^{#1}$ Physique Math\'ematique des Interactions Fondamentales \& International Solvay Institutes, Universit\'e Libre de Bruxelles, Campus Plaine - CP 231, 1050 Bruxelles, Belgium
			\end{center}
		\end{minipage}
		}
	}
\newcommand{\uiuc}[1]{
	\centerline{
		\begin{minipage}[c]{0.7\textwidth}
			\begin{center}
			${}^{#1}$ Illinois Center for Advanced Studies of the Universe \& Department of Physics,\\ 
			University of Illinois, 1110 West Green St., Urbana IL 61801, U.S.A.
			\end{center}
		\end{minipage}
		}
	}

\begin{document}

\title{\thistitle}
\author{
	Luca Ciambelli$^{a}$ and Robert G. Leigh$^{b}$
	\\
	\\
	{\small \emph{\ulb{a}}} \\ \\
	{\small \emph{\uiuc{b}}}
	\\
	}
\date{}
\maketitle
\vspace{-1ex}
\begin{abstract}
\vspace{0.6cm}
Conserved charges in theories with gauge symmetries are supported on codimension-2 surfaces in the bulk spacetime. 
It has recently been suggested that various classical formulations of gravity dynamics display different symmetries, and paying attention to the maximal such symmetry could have important consequences to further elucidate the quantization of gravity. After establishing an algebraic off-shell derivation of the maximal closed subalgebra of the full bulk diffeomorphisms in the presence of an isolated corner, we show how to geometrically describe the latter and its embedding in spacetime, without constraining the geometry away from the corner, such as by assuming a foliation. The analysis encompasses arbitrary embedded surfaces, of generic codimensions $k$. The resulting corner algebra ${\cal A}_k$, calling $S$ the embedded surface and $M$ the bulk, is that of the group $(Diff(S)\ltimes GL(k,\RR))\ltimes \RR^k$. This result is independent of any dynamics or pseudo-Riemannian structure in the bulk. We then evaluate the Noether charges of ${\cal A}_2$ for Einstein-Hilbert dynamics and show that the Noether charge algebra gives a representation of the algebra  ${\cal A}_2$, for finite proper distance corners in the bulk, while all other charges associated with $Diff(M)$ vanish.
\end{abstract}

\newpage


\setcounter{footnote}{0}
\renewcommand{\thefootnote}{\arabic{footnote}}

\section{Introduction}

In the Hamiltonian \cite{Arnowitt:1962hi, Regge:1974zd, Henneaux:1985tv, Brown:1986ed, Brown:1986nw} or covariant phase space formulation \cite{Kijowski:1976ze, Crnkovic:1986ex, Lee:1990nz, ASHTEKAR1991417, Wald:1993nt, Iyer:1994ys, Wald:1999wa} of physical theories, focus is often put on codimension-1 hypersurfaces in spacetime, upon which one imagines specifying initial field configurations. Nevertheless, it has been known since the time of Noether's seminal paper \cite{doi:10.1080/00411457108231446}, and perhaps before in some contexts, that conserved gauge charges have support only in codimension-2 in spacetime. This arises for example in electrodynamics, via Gauss' law, in which the codimension-2 surfaces are usually taken to be spheres surrounding charged matter.
Such charges are also the objects of interest in the Lagrangian reducibility parameter method initiated in \cite{Barnich:1995ap, Barnich:2000zw} and further developed in \cite{Barnich:2001jy, Barnich:2007bf}. We will follow more recent literature and refer generically to spacelike codimension-$2$ surfaces, upon which charges may be defined, as \hlt{corners}. The interest in corners goes well beyond the classical dynamics of gauge theories and gravity; indeed in addition to their relevance to conserved charges, examples of corners appear as entanglement cuts in quantum field theories, and as horizons and celestial spheres, and thus scattering amplitudes.  As first discussed in \cite{Freidel:2015gpa, Donnelly:2016auv}, it is therefore expected that corners (and symmetries thereof) should play a prominent role in the quantization of any gauge theories, gravity included, a concept that was also recently emphasized in \cite{Freidel:2020xyx, Freidel:2020svx,  Freidel:2020ayo}. In the case of gravity at least, one has in mind here a sort of generalized holographic structure, in which the basic building blocks of a quantum theory would be defined on corners, and ``bulk'' geometries in which they are thought to be immersed arising in some semi-classical limit. In that context, it is indeed of the utmost importance to organize the physics of the corners in a clear geometric fashion, with the symmetries (whether they be spontaneously broken or not) and their representation theory clearly elucidated.

In the absence of any specified geometric structure defining a particular locus of points in a manifold $M$ of dimension $d$, the group of transformations allowed on $M$ in gravity is its group of diffeomorphisms $Diff(M)$, and it is purely a redundancy of the theory. This is the basis of any geometric theory. In general relativity, it is traditional to specify a pseudo-Riemannian structure on the bulk, and one often considers particular classes of metrics or invokes simplifying assumptions. But foremost, one almost always chooses a specific gauge (or choice of coordinates). Given recent developments on the relevance of edges and corners \cite{Donnelly:2014fua, Donnelly:2016auv, Speranza:2017gxd, Freidel:2020xyx, Freidel:2020svx,  Freidel:2020ayo, Donnelly:2020xgu}, it has become increasingly clear that gauge-fixing in an empty spacetime is quite different than gauge-fixing in a spacetime with specified substructures, such as corners, and indeed one expects to find in general degrees of freedom that have support on such substructures that would not be considered physical in the bulk. This is by no means a new or controversial idea, as it is a well-known feature, for example, of topological gauge theories. Indeed, in the context of entanglement for example, this is the root cause of the spatial non-factorizability of Hilbert space in gauge theories. In many applications,  corners can be regarded as isolated submanifolds\footnote{Throughout the paper, our analysis will be essentially local and will concern mostly properties of tangent bundles rather than the corresponding base spaces themselves. Thus, one can think in terms of the submanifolds under discussion as being immersions, but nevertheless we will often use the language of embeddings for clarity.} in $M$, and the local geometry in a neighbourhood of a corner serves to describe the embedding without specifying bulk dynamics. One of the purposes of this paper is to describe the relevant geometry in a neighbourhood of a corner in a general way without specifying additional structure such as foliations. Another purpose is to separate the aspects that pertain to a particular dynamical theory from those that are purely geometric (or, therefore ``kinematic''). In particular, in the context of a theory involving two derivatives of a metric on $M$, without specifying particular dynamics or a corresponding classical solution, one expects to be able to constrain the local geometry up to and including a first derivative. 
Constraining the geometry further in this context is equivalent to introducing dynamics. 

The presence of a corner then removes part of the bulk redundancies, making them true physical symmetries, with non-vanishing surface charges. These symmetries are the so-called corner symmetries,\footnote{We use the word symmetry here to denote an underlying group of transformations of interest. It should be noted that we are not using it in the sense of a property of some action or of some particular background.} and to our knowledge, they appeared first in the literature in Ref. \cite{Regge:1974zd}. 
Indeed, given a manifold $S$ of dimension $n=d-2$ embedded or immersed in $M$, it is natural to regard data (an induced metric, connections, etc.) on the manifold as background fields whose components respond to changes of coordinates on $S$. From this point of view, we regard $Diff(S)$ as a background symmetry. This background symmetry can be expected to be extended, corresponding to freedom in how $S$ is embedded in $M$.\footnote{Often in the literature, a related problem is addressed where one is interested in fixing some geometric structures on $S$ and then considering the transformations that preserve this structure.} 
 The embedding of $S$ in a manifold $M$ can be specified as a map $\phi:S\to M$, although we are primarily interested in the corresponding injection $d\phi:TS\to TM$. The latter map can be thought of in terms of a (generally non-integrable) screening distribution (that is, a sub-bundle of $TM$), requiring that the normal forms defining the distribution pull back to zero on $S$. We will make use of this construction in the present paper, and we note that we have recently made use of similar tools in studying local Weyl invariance in holography \cite{Ciambelli:2019bzz} and in constructing the Carroll fibre bundle \cite{Ciambelli:2019lap} on null geometries. The construction makes manifest additional symmetries, contained in $Diff(M)$, that act naturally on the embedding. Indeed, one of our first tasks in this paper will be to show that, in the more general context of codimension-$k$ (i.e., $d=n+k$), there is a maximal closed subalgebra of the algebra $\dM$  that organizes the local geometry, which we call the \hlt{maximal embedding algebra}. 

One of the central concepts in this construction is the relevance of a so-called Ehresmann connection which appears in a suitable parameterization of a metric on $M$ in the presence of a corner. 
Indeed we will parameterize the bulk metric using a generalized version of the so-called Randers-Papapetrou form \cite{PhysRev.59.195, Papapetrou:1966zz, Gibbons:2008zi}. This is an alternative to the more commonly utilized Zermelo parameterization \cite{https://doi.org/10.1002/zamm.19310110205}, which as we will explain, implicitly assumes a codimension-2 foliation. 
The Randers-Papapetrou form is, in contrast, appropriate in the context of an isolated codimension-$k$ surface. This fact is borne out by an analysis and interpretation of the symmetries (corresponding to the aforementioned maximal subalgebra of $\dM$) that act simply on the fields appearing in the parameterization. 
This phenomenon is analogous to the appearance of $\mathfrak{bms}$ algebras on Carroll structures  \cite{Ciambelli:2019lap}. 
As we will see, the interplay of algebraic and geometric features permeates our analysis in a coherent and simple way. 

There is another reason for utilizing a generalized Randers-Papapetrou parameterization. Since the codimension-$2$ corner is the support for surface charges, we will take it to be spacelike, with any notion of time corresponding to a normal extrinsic direction. Correspondingly, the normal bundle has signature $(-1,1)$. By reversing the logic, while Zermelo foliates corners, Randers-Papapetrou foliates the normal bundle, described geometrically, for example, recently in \cite{Speranza:2019hkr}. 
This is a geometric way to describe the fact that there is a $2$-dimensional lightcone for each point on $S$, a picture that is certainly relevant for example in the description of scattering amplitudes \cite{Kapec:2014opa, Campiglia:2015yka, Pasterski:2016qvg, Strominger:2017zoo, Donnay:2020guq} on asymptotically flat geometries where the corner is the celestial sphere: gravitons or photons freely propagate along straight lines on $M$, hitting $S$ at a point. To some extent this resonates with the original discussion that led Penrose to the definition of the Penrose-Carter diagrams \cite{Penrose:1962ij, Carter:1966zza, Hawking:1973uf}. Furthermore, it is in line with the logic used in methods in General Relativity to describe Einstein solutions, like the Newman-Penrose \cite{Newman:1961qr} or Geroch-Held-Penrose \cite{Geroch:1973am} formalisms.


The paper is organized as follows. In Section \ref{sec2.1}, we describe the construction of a codimension-$k$ distribution in $TM$ as the kernel of a set of 1-forms. In suitable adapted coordinates, these 1-forms in general include a set of functions that will have an interpretation as components of an Ehresmann connection. The condition that these forms pull back to zero on $S$ is shown to correspond to a flatness condition on the pullback connection, which is weaker than an integrability condition. In Section \ref{sec:maxembalg}, we establish the closure of a subalgebra of $\dM$ that corresponds to symmetries of a codimension-$k$ submanifold. The maximal such algebra is the algebra ${\cal A}_k=\Big(\mathfrak{diff}(S)\loplus \mathfrak{gl}(k,\RR)\Big)\loplus \RR^k$ of the group $\Big(Diff(S)\ltimes GL(k,\RR)\Big)\ltimes \RR^k$. By construction, the subalgebra is realized in the bulk through the Lie bracket of certain vector fields on $M$, some of which vanish on $S$. In Section \ref{sec23}, we show that the connection appearing in the parameterization of the normal forms defining the distribution give rise, through an expansion near $S$, to a collection of fields defined on $S$ that transform non-linearly (that is, as gauge fields) with respect to $\RR^k$ and $GL(k,\RR)$. (In a later section, we will show that a connection for $Diff(S)$ arises by induction from the bulk Levi-Civita connection upon introduction of a metric on $M$). In this way, the full symmetry ${\cal A}_k$ can be thought to be implemented intrinsically on the isolated surface $S$. 

In Section \ref{sec:metric}, we consider the introduction of a metric on $M$ in the Randers-Papapetrou form, making direct use of the normal forms discussed earlier. We demonstrate that the action of ${\cal A}_k$ on the components of the metric have a simple geometric interpretation.
In Section \ref{sec:dynamics}, we consider the Noether charge in the Einstein-Hilbert theory. In this context we are interested in codimension-2. We show that the non-zero Noether charges correspond to a subset of the vector fields generating the ${\cal A}_2$ diffeomorphisms. This is a generalization of the results of \cite{Donnelly:2020xgu}; indeed we find that the full ${\cal A}_2$ algebra is not realized on the Einstein-Hilbert phase space, but instead a subalgebra, $\tilde{\cal A}_2=\Big(\mathfrak{diff}(S)\loplus \mathfrak{sl}(2,\RR)\Big)\loplus \RR^2$, at least in the case of a generic surface $S$ (that is, asymptotic embeddings should be considered separately, and we will return to them in a separate publication). The $\RR^2$ factor corresponds to local translations of the surface (i.e., supertranslations), a symmetry which is spontaneously broken by any fixed surface $S$. We show that the  algebra of the Noether charges closes on the phase space (that is, gives rise to a consistent set of brackets) if we carefully take into account not only variations of the metric but of the embedding itself. The variation of the embedding is required to close the algebra in the presence of translations, and should be thought of as corresponding to including the embedding as part of the field content of the theory. We stress again that we perform this analysis for finite proper distance corners and in the context where we do not impose specific gauge conditions on fields or boundary conditions on the corner. We show how the results are extended to the case in which such gauge conditions are imposed, by taking into account the corresponding field-dependence of the vector fields generating the charges. This effect does not modify the charges themselves, but it does play a role in the algebra of the charges, which then involves a modified bracket. 
In Section \ref{sec:DGGP}, we consider a specific example of the symmetries of a black hole horizon, and we show in detail how our formalism compares with the treatment in the literature, as e.g., \cite{Donnay:2015abr}.
Finally, in Section \ref{sec:remarks}, we conclude with comments and direction for future research. 
In Appendix \ref{AppB}, we discuss some details of how the bulk Levi-Civita connection induces a connection on the corner, while in Appendix \ref{AppA} we give an explicit computation of the change of embedding and its active versus passive interpretation.

\section{Invariant Description of Embedded Surfaces}\label{sec:distr}

In this section we study $d$-dimensional manifolds $M$ admitting isolated surfaces $S$.
We begin the discussion with the geometric aspects of the embedding, and later focus on  algebraic aspects. 
We emphasize that everything in this section is  off-shell in that we will not yet introduce a metric and thus the discussion is independent of any particular dynamics.

\subsection{Geometric Structure}\label{sec2.1}

Consider an isolated codimension-$k$ surface $S$, embedded in $M$
\beq\label{embedding}
\phi:S\to M.
\eeq
In the case $k=2$, this will be a corner,  although for the sake of discussion we leave $k$ arbitrary in this section.
By using the term isolated, we mean that we are not assuming that $M$ is foliated, even locally, by such surfaces. Of course this does not preclude a foliation, but we are not bound by its existence. What we are assuming in this paper is that $S$ is embedded (or somewhat more generally, immersed) in $M$, and a choice of the map $\phi$ specifies this embedding. Where useful, we will denote a choice of coordinates on a local patch of $S$ as $\{\sigma^\alpha, \alpha=1,...,n\}$, with $n=d-k$. Given local coordinates $y^M$ on $M$, the embedding $\phi$ can be expressed by giving $y^M=y^M(\sigma)$. A particularly nice choice that we will often refer to is where we adapt the coordinates such that the embedding corresponds locally to the vanishing of $k$ functions, $u^a\in C^\infty(M)$, and the embedded surface has $u^a(\sigma)=0,\forall a=1,...,k$. Furthermore, one often chooses the coordinates $y^M=(u^a,x^i)$, with the embedding further given by $x^i(\sigma)=\delta^i_\alpha\sigma^\alpha$. We will refer to such an embedding as $\phi_0$, with $\phi_0:y_0^M(\sigma)=(0,\delta^i_\alpha\sigma^\alpha)$.

It is natural to introduce a split of the tangent bundle $TM$ in a vertical rank-$k$ sub-bundle $V$ and a complementary rank-$n$ horizontal sub-bundle $H$. The embedding $\phi$ is ``adapted'' to such a structure if
\beq\label{conds}
\phi^*(H^*)=T^*S,\qquad \phi^*(V^*)=0,
\eeq
where $\phi^*$ is the pull-back of $\phi$ and $H^*,V^*$ are the dual bundles of $H$ and $V$. 
The split of $TM$ in $H$ and $V$ is a local statement, preserved only under a subset of the diffeomorphisms in $M$, which we will emphasize below. 
By introducing $k$ 1-forms $n^a\in T^*M$ (regarded as in $V^*$), we define $H$ to be the distribution 
\beq\label{defgendis}
C_k(n^a)\equiv ker(\{n^a\})=\left\{ \un\mX\in TM\Big| n^a(\un\mX)=0,\forall a=1,..,k\right\}.
\eeq
An adapted embedding $\phi$ is then subject to the condition \eqref{conds} that the normal forms $n^a$ pull back to zero on $S$, $\phi^*n^a=0,$ that is
\beq\label{pullbackn}
\phi^*n^a(\un X)=0=n^a(\phi_*\un X),\qquad \forall \un X\in TS.
\eeq
If we specify $n^a$, then we should interpret this as a condition on the embedding.

Let us now introduce local coordinates in $M$ that cover a neighbourhood of an open set in the embedded surface. As described above, we will refer to a choice of local coordinates on $M$ as $y^M=(u^a,x^i)$, with $a=0,..,k-1$ and $i=1,...,n$.  
Up to normalization,\footnote{Note that the definition of $C_k(n^a)$ does not constrain the normalization of the $n^a$, and so multiplying the $n^a$ by functions on $M$ does not modify the distribution. We also assume that, at least near the embedded surface the 1-forms $n^a$ are nowhere vanishing and linearly independent, such that the tangent planes to the image of $S$ are everywhere of the same dimension.} we can write the forms $n^a$ as 
\beq\label{defthens}
n^a\equiv du^a-a^a_i(u,x)dx^i.
\eeq
Given such a distribution, there is a notion of integrability which (by the Frobenius theorem) would imply that $M$ is foliated. Since we are generally interested in just an isolated embedded surface, we will make no such assumption, beyond the pullback condition \eqref{pullbackn}. A simple example of a foliation would be obtained by taking $a^a_i(u,x)$ to identically vanish, in which case the leaves of the foliation are just the level surfaces of the $u^a$. The reader familiar with Ehresmann connections may anticipate that $a^a_i(u,x)$ may be interpreted in such terms, representing the ambiguity in lifting vectors in $S$ to a horizontal sub-bundle of $TM$. We will establish in what sense such a picture pertains, but we emphasize that the construction should be thought of as valid in a neighbourhood of the embedded surface, and certainly not far away. In this sense, we organize tensorial objects that can be thought of as intrinsic to $S$; extending into $M$ further should be thought of in terms of introducing dynamics.

To further understand the pull-back condition, for an  embedding $y^M(\sigma)=(u^a(\sigma),x^i(\sigma))$, we find that the pull back of the bulk forms $n^a$ are given explicitly by
\beq
\phi^*n^a = (\pa_\alpha u^a(\sigma)-a^a_i(u(\sigma),x(\sigma))\pa_\alpha x^i(\sigma))d\sigma^\alpha,\qquad 
\phi^*dx^i=\pa_\alpha x^i(\sigma) d\sigma^\alpha
\eeq
and so $\phi^*n^a =0$ gives 
\beq\label{transgfpuregauge}
\pa_\alpha u^a(\sigma)=\pa_\alpha x^i(\sigma) a^a_i(u(\sigma),x(\sigma))\equiv a^a_\alpha(\sigma).
\eeq
One can understand this as a differential equation for the embedding $(u^a(\sigma),x^i(\sigma))$. Equivalently, we
 interpret this to mean that $a_i^a(u,x)dx^i$  pulls back under $\phi$ to $a^a_\alpha(\sigma)d\sigma^\alpha$ and this must be simply $\pa_\alpha u^a(\sigma)d\sigma^\alpha$.  So one can interpret $a^a_\alpha(\sigma)$ as a (flat) connection for the normal translation 
group\footnote{We make this identification because a non-trivial embedding $\phi$ can be thought of as corresponding to a surface described by $u^a(\sigma)$, that is, a local translation with respect to the ``trivial" embedding $\phi_0$. So this by definition corresponds to a diffeomorphism that does not preserve (i.e., moves) the surface.} 
$\RR^k$, which is spontaneously broken by the presence of $S$, with the $u^a(\sigma)$ the corresponding Goldstone modes. Note that eq. \eqref{transgfpuregauge} is not a restriction on $a_i^a(u,x)$, but on its pullback once a specific embedding is chosen. What eq. \eqref{transgfpuregauge} expresses is the fact that the assumption that $S$ is embedded leads to the flatness of the connection $a^a_\alpha$ on $S$, with the choice of gauge for this flat connection determined by the embedding. In the specific embedding $\phi_0$ we have $a^a_\alpha=0$, but nevertheless the $u$-derivatives of $a^a_i(u,x)$ will give rise to non-trivial intrinsically defined quantities on $S$ that are of physical interest.

Given the coordinatization \eqref{defthens}, it is straightforward to solve \eqref{defgendis}. One finds that the distribution $H$ is spanned by vector fields $\un D_i$,
\beq
C_k(n^a)=span\big\{ \un D_i=\un\pa_i+a_i^a(u,x)\un\pa_a\big\}.
\eeq
Therefore, the $a_i^a(u,x)$ have an interpretation in terms of the components (here in a local trivialization) of an Ehresmann connection, necessary to lift the ambiguity in the definition of the distribution \eqref{defgendis} as the kernel of the vertical one forms $n^a$. Notice that $\pa_\alpha x^i (\un\pa_i+a_i^a(u(\sigma),x(\sigma))\un\pa_a)=\un\pa_\alpha$ by the chain rule given \eqref{transgfpuregauge}, and thus $\un D_i$ is the push-forward of $\un\pa_\alpha$. Consider the Lie bracket 
\beq\label{curv}
\big[ \un D_i,\un D_j\big]= (D_ia_j^a-D_ja_i^a)\un\pa_a\equiv f_{ij}^a(u,x)\un\pa_a.
\eeq
If we interpret $f_{ij}^a(u,x)$ as the components of a $2$-form and pull this back to $S$, we find just $f^a_{\alpha\beta}(\sigma)=\pa_\alpha a_\beta^a(\sigma)-\pa_\beta a_\alpha^a(\sigma)$ which vanishes given \eqref{transgfpuregauge}. So indeed, one can view the pullback condition as a flatness condition for the translations. Note that this is a statement in $S$ that therefore does not preclude the distribution $H$ from being  non-integrable in $M$. As we will discuss in Section \ref{sec23}, there is a subset of diffeomorphisms of the form $u'(u,x)$ and $x'(x)$
under which the field $a^a_i$ transforms as an Ehresmann connection. While this subset of diffeomorphisms preserves 
the split of the tangent bundle, it does not preserve the isolated surface $S$. As we will now show, the latter is preserved only under an even smaller subclass of diffeomorphisms included in those just mentioned. 

\subsection{Maximal Embedding Algebra}\label{sec:maxembalg}

As we have just seen, a natural choice of embedding is $u^a(\sigma)=0, x^i(\sigma)=\delta^i_\alpha\sigma^\alpha$; more generally we can regard this as a specific choice of gauge for the (spontaneously broken) normal translations, and our parameterization given in the last section emphasizes the corresponding pure gauge connection. We interpret the coordinates $(u^a,x^i)$ as adapted to this trivial embedding, and generally consider embeddings that are `close to' the trivial embedding.
As such, we will include in this discussion transformations that correspond to infinitesimal departures from this gauge, which are precisely the local translations $\mathbb{R}^k$. Indeed, this symmetry plays an interesting role in certain applications, as we will uncover. Given a vector field $\un\xi=\xi^i(u,x)\un\pa_i+\xi^a(u,x)\un\pa_a$ on $M$, we  expand the components near $u^a=0$,\footnote{While it is possible that non-analytic contributions occur in these expansions, we expect it to be always possible to probe a neighbourhood of a point to first order analytically.}
\beqn\label{exp}
\xi^b(u,x)&=&\sum_{n=0} \frac{1}{n!}u^{a_1}...u^{a_n} \xi_{(n)}^b{}_{a_1...a_n}(x)=\xi_{(0)}^b(x)+\xi_{(1)}{}^b{}_{a_1}(x)u^{a_1}+{1\over 2}\xi_{(2)}{}^b{}_{a_1a_2}(x)u^{a_1}u^{a_2}+\dots
\\
\xi^i(u,x)&=&\sum_{n=0} \frac{1}{n!}u^{a_1}...u^{a_n} \xi_{(n)}^i{}_{a_1...a_n}(x)=\xi_{(0)}^i(x)+\xi_{(1)}^i{}_{a_1}(x)u^{a_1}+{1\over 2}\xi_{(2)}^i{}_{a_1a_2}(x)u^{a_1}u^{a_2}+\dots
\eeqn
Nevertheless, we are not interested here in studying under what conditions these expansions are possible to all orders of $u^a$ because, as we will see shortly, there is a natural truncation at first order.\footnote{Often in the literature, one encounters vector fields that have further subleading, field dependent, terms. These occur because in such contexts one is imposing conditions on fields and requiring that the transformations preserve such conditions. Here we are imposing no conditions and thus it is consistent to have such a truncation. In any case, the leading terms in the vector fields do determine the gauge charges and their algebra, at least in the case of finite proper distance corners that we consider in this paper.  Later in the paper, in the context of Noether charges in the Einstein-Hilbert theory with support on general codimension-2 corners, we will return to this issue and understand what simple modifications must be made.} 
We then consider the terms up to first order in the Lie bracket of two such vector fields,
\beqn\label{LieM}
\big[\un\xi_1,\un\xi_2\big]
&=&
\Big[\colr{\un{\hat\xi}_{(0)1}}(\colb{\xi_{(0)2}^b})-\colg{\xi_{(1)1}{}^b{}_a}\colb{\xi_{(0)2}^a}
+\colg{\xi_{(1)2}{}^b{}_a}\colb{\xi_{(0)1}^a}-\colr{\un{\hat\xi}_{(0)2}}(\colb{\xi_{(0)1}^b})\Big]\un\pa_b
\\&&\nonumber
+\Big[\xi_{(1)2}^j{}_a\colb{\xi_{(0)1}^a}-\xi_{(1)1}^j{}_a\colb{\xi_{(0)2}^a}
+\colr{\big[\un{\hat\xi}_{(0)1},\un{\hat\xi}_{(0)2}\big]^j}\Big]\un\pa_j
+u^c\Big(
\Big[-\colg{\big[\xi_{(1)1},\xi_{(1)2}\big]^b{}_c}
+\xi_{(2)2}{}^b{}_{ca}\colb{\xi_{(0)1}^a}
\\&&\nonumber-\xi_{(2)1}{}^b{}_{ca}\colb{\xi_{(0)2}^a}
+\colr{\un{\hat\xi}_{(0)1}}(\colg{\xi_{(1)2}{}^b{}_c})
-\colr{\un{\hat\xi}_{(0)2}}(\colg{\xi_{(1)1}{}^b{}_c})
+\un{\hat\xi}_{(1)1c}(\colb{\xi_{(0)2}^b})
-\un{\hat\xi}_{(1)2c}(\colb{\xi_{(0)1}^b})
\Big]\un\pa_b
\\&&\nonumber
+\Big[
\xi_{(1)2}^j{}_{a}\colg{\xi_{(1)1}{}^a{}_c}
-\xi_{(1)1}^j{}_{a}\colg{\xi_{(1)2}{}^a{}_c}
+\big[\un{\hat\xi}_{(1)1c},\colr{\un{\hat\xi}_{(0)2}}\big]^j
-\big[\un{\hat\xi}_{(1)2c},\colr{\un{\hat\xi}_{(0)1}}\big]^j
+\xi_{(2)2}^j{}_{ca}\colb{\xi_{(0)1}^a}
-\xi_{(2)1}^j{}_{ca}\colb{\xi_{(0)2}^a}
\Big]\un\pa_j
\Big)
+...
\eeqn
where by hats we denote vector fields of the form $\un{\hat\xi}=\xi^i\un\pa_i$. We see that there is a  consistent truncation of the
 full $\dM$ algebra generated by vector fields of the form
\beq\label{subalgebragens}
\un{\xi}=\xi_{(0)}^k(x)\un\pa_k +\Big(\xi_{(0)}^a(x)+u^b\xi_{(1)}{}^a{}_b(x)\Big)\un\pa_a.
\eeq
Indeed, these vector fields
satisfy the closed algebra
\beqn\label{LieMsub}
\big[\un{\xi}_1,\un{\xi}_2\big]
&=&
\big[\un{\hat\xi}_{(0)1},\un{\hat\xi}_{(0)2}\big]^j\un\pa_j
\\&&\nonumber
+\Big[
\un{\hat\xi}_{(0)1}(\xi_{(0)2}^b)
-\un{\hat\xi}_{(0)2}(\xi_{(0)1}^b)
-\xi_{(1)1}{}^b{}_a\xi_{(0)2}^a
+\xi_{(1)2}{}^b{}_a\xi_{(0)1}^a
\Big]\un\pa_b
\\&&\nonumber 
+u^c
\Big[-\big[\xi_{(1)1},\xi_{(1)2}\big]^b{}_c
+\un{\hat\xi}_{(0)1}(\xi_{(1)2}{}^b{}_c)
-\un{\hat\xi}_{(0)2}(\xi_{(1)1}{}^b{}_c)
\Big]\un\pa_b.
\eeqn
To interpret this algebra from the point of view of a corner $S$, it is particularly simple to consider the trivial embedding $\phi_0$, in terms of which the first line can be interpreted as simply the push forward of the Lie bracket on $S$; as such, we will refer to this as the algebra $\dS$, which is itself a subalgebra of the full bulk algebra. On the third line, we have the commutator of matrices of the form $\xi_{(1)}{}^a{}_b(x)$, giving a $\mathfrak{gl}(k,\RR)$ algebra.
In addition, the third line contains terms corresponding to the action of $\dS$ on $\mathfrak{gl}(k,\RR)$. Thus, we have a $\dS\loplus \mathfrak{gl}(k,\RR)$ subalgebra, given that the second line vanishes with $\xi^a_{(0)}$. Without that restriction the second line contains both $\dS$ and $\mathfrak{gl}(k,\RR)$ acting on $\xi^a_{(0)}(x)$, which are the vector fields generating the Abelian algebra $\RR^{k}$ corresponding to translations in the normal directions discussed above (they do not preserve the $u^a(\sigma)=0$ gauge). Thus the full algebra is ${\cal A}_k=\Big(\mathfrak{diff}(S)\loplus \mathfrak{gl}(k,\RR)\Big)\loplus \RR^k$, which we will generally refer to as the maximal embedding algebra. This is the algebra of the group $\Big(Diff(S)\ltimes GL(k,\RR)\Big)\ltimes \RR^k$. The generators are
\beq\label{maximo}
\Big(\underbrace{Diff(S)}_\text{$\xi^j_{(0)}$}\ltimes \underbrace{GL(k,\RR)}_\text{$\xi_{(1)}{}^a{}_b$}\Big)\ltimes \underbrace{\RR^{k}}_\text{$\xi_{(0)}^a$}.
\eeq
To summarize, we have
\beqn\label{LieMsubexpl}
\underbrace{\big[\un{\xi}_1,\un{\xi}_2\big]}_{{\cal A}_k}
&=&
\underbrace{\big[\un{\hat\xi}_{(0)1},\un{\hat\xi}_{(0)2}\big]^j\un\pa_j}_\text{$\dS$}
\\&&
+\Big[
\underbrace{\un{\hat\xi}_{(0)1}(\xi_{(0)2}^b)
-\un{\hat\xi}_{(0)2}(\xi_{(0)1}^b)}_\text{$\dS$ acts on $\RR^k$}
+\underbrace{\xi_{(1)2}{}^b{}_a\xi_{(0)1}^a
-\xi_{(1)1}{}^b{}_a\xi_{(0)2}^a}_\text{$\mathfrak{gl}(k,\RR)$ acts on $\RR^k$}
\Big]\un\pa_b
\\&&
+u^c
\Big[-\underbrace{\big[\xi_{(1)1},\xi_{(1)2}\big]^b{}_c}_\text{$\mathfrak{gl}(k,\RR)$}
+\underbrace{\un{\hat\xi}_{(0)1}(\xi_{(1)2}{}^b{}_c)
-\un{\hat\xi}_{(0)2}(\xi_{(1)1}{}^b{}_c)}_\text{$\dS$ acts on $\mathfrak{gl}(k,\RR)$}
\Big]\un\pa_b.
\eeqn
This algebra for codimension-2, or at least the $\dS\loplus \mathfrak{sl}(2,\RR)$ subalgebra, is the one studied recently in \cite{Donnelly:2020xgu}, whose importance was appreciated in \cite{Donnelly:2016auv}. In that context, this subalgebra is the one that becomes physical on the codimension-2 corner $S$, while the rest of the $\dM$ transformations keep generating spacetime redundancies.  It is remarkable that with this simple derivation we have access to such an apparently deep feature of the theory. We will see in the following sections that a connection for this full symmetry is contained in the $a_i^a(u,x)$ field together with the induced connection on $S$ (given a metric in the bulk, which we will introduce in Section \ref{sec:metric}). 

Before moving to such a study, we note that ${\cal A}_k$, in the presence of an isolated embedded surface, is the maximal {\it closed} subalgebra of $\dM$ admitting a finite expansion in powers of $u^a$. Indeed one finds that including any further terms beyond those in \eqref{subalgebragens} leads to an infinite algebra. For instance, one can consider transformations of the type $u'(u,x)$ and $x'(x)$. Including all powers of $u^a$, this algebra is smaller than $\dM$ and closes. The biggest consistent truncation of this algebra to finite powers of $u^a$ is then the algebra ${\cal A}_k$. As already stressed, in this work we are not assuming a codimension-$k$ foliation in $M$. That is, we are considering an isolated surface rather than a family of them. Therefore, transformations of the type $u'(u,x),x'(x)$ have a clear geometric interpretation as symmetries on the single surface ($x'(x)$) and generic transformations of the extrinsic space, preserving 
the split of the tangent bundle. On the other hand, had we assumed a family of surfaces foliating the bulk, we would have been naturally interested in diffeomorphisms acting differently on each leaf, i.e., transformations of the form $x'(u,x)$, together with $u'(u)$, that avoid mixing the coordinates labeling the leaves with the intrinsic ones. We see therefore that, although there is nothing fundamentally wrong in introducing a family of embedded surfaces in $M$, not only  it is not the most general setup to investigate but also the symmetries we found, being a particular subset of $u'(u,x),x'(x)$, are an indication that the geometric structure established here is most suitable.

We should also note that in the case $k=2$, the algebra ${\cal A}_2$ is not necessarily completely realized on the phase space of a given dynamical theory. This might happen if there are simply no degrees of freedom on $S$ that are charged under some subset of the symmetry.
Although we will not explore all possibilities in the current paper, we note that this may occur depending on the details of a specific corner's embedding, because the symmetries discussed here are linearly realized via coordinates in the ambient space (that is, the bulk).

In the special case $k=1$, we have the group $\Big(Diff(S)\ltimes \RR\Big)\ltimes \RR$. In the context of asymptotically $AdS$ spacetimes with $S$ the conformal boundary, the first $\RR$ factor is the Weyl diffeomorphism discussed in \cite{Ciambelli:2019bzz}. The second $\RR$ is a simple translation of the radial coordinate $z$, which is usually considered fixed, the boundary taken at $z=0$ (more generally, the translation can be taken to induce a spacetime-varying cutoff). Another application of the codimension-$1$ case occurs in the study of corners of certain asymptotic realizations of null structures. In that context, the first $\RR$ factor is not independent, being related to the divergence of the $Diff(S)$ generators, an example of a field-dependent transformation introduced to preserve a particular structure. The second $\RR$ factor is free and corresponds to the supertranslations, giving rise to the {\it generalized} $BMS$ group occurring in the asymptotic analysis \cite{Campiglia:2014yka, Compere:2018ylh}.\footnote{The original $BMS$ analysis \cite{doi:10.1098/rspa.1962.0161, doi:10.1098/rspa.1962.0206} restricted $\dS$ to the globally well-defined conformal Killing isometries of $S$. The extension to local, not necessarily invertible, holomorphic mappings has been considered in \cite{Barnich:2010eb}, leading to the extended $BMS$ group. The embedding of $BMS$ in $AdS$ and $dS$ spaces has been discussed in \cite{Compere:2019bua}.} 
We therefore see that it is important to keep track of the normal translation symmetry, as it plays a crucial role in specific realizations of our setup.
The generalized $BMS$ group has been further enlarged recently to include Weyl in \cite{Freidel:2021yqe}; the resulting group there is $\Big(Diff(S)\ltimes \RR\Big)\ltimes \RR$, where the Weyl $\RR$ factor has been disentangled from $Diff(S)$. This group is also discussed in \cite{Adami:2020ugu} for $2$- and $3$-dimensional bulks.

\subsection{Ehresmann Connection}\label{sec23}

Now let us return to the Ehresmann connection that we first discussed in Section \ref{sec2.1}, encoded in the definition of the distribution $C_k(n^a)$. The distribution defines a horizontal sub-bundle inside $TM$. Although one can perform arbitrary diffeomorphisms in $M$, only a subset of them preserves the 
the local split of $TM$ into $H$ and $V$. As we already discussed, these are the finite transformations of the form $u'(u,x)$ and $x'(x)$.  Under them, we require that the vertical sub-bundle is preserved, that is
\beq
 n'^a(u'(u,x),x'(x))=J^a{}_bn^b(u,x), \quad \text{with}\quad J^a{}_b={\pa u'^a\over\pa u^b}.
\eeq
This requirement implies that the $a^a_i$ transform non-linearly as
\beq\label{aprime}
a'^a_i(u'(u,x),x'(x))=J^{-1i}{}_j(J^a{}_b a^b_i+J^a{}_i), \quad \text{with}\quad J^i{}_j={\pa x'^i\over\pa x^j}, \ J^a{}_i={\pa u'^a\over \pa x^i},
\eeq
where we have used that the Jacobian for this finite diffeomorphism is upper triangular, $J^i{}_a=0$. It is the last piece in the transformation of $a^a_i$ in \eqref{aprime} that  confers to it the status of Ehresmann connection under the group preserving the split.

We now expand infinitesimally the transformation \eqref{aprime}. To do so, we write $u'^a=u^a-\xi^a$ and $x'^i=x^i-\xi^i$ keeping both $\xi^a$ and $\xi^i$ arbitrary, and as usual we expand the left-hand side near $(u,x)$ defining $a'^a_i(u,x)=a^a_i(u,x)+\delta_{\un\xi}a^a_i(u,x)$. We then obtain
\beq\label{dela}
\delta_{\un\xi}a^a_i(u,x)=
a^a_j\pa_i\xi^j
+\xi^j\pa_j a^a_i
+\xi^b\pa_b a^a_i
-a^b_i\pa_b \xi^a
-\pa_i\xi^a.
\eeq
Due to the semi-direct structure in \eqref{maximo}, the finite version of \eqref{subalgebragens} falls inside the class of transformations studied here. Therefore, although so far more general, we can now restrict our attention to the embedding symmetries generated by \eqref{subalgebragens}. 
Expanding the bulk field $a_i^a(u,x)$ near $u^a=0$ as
\beq
a_i^a(u,x)=a^{(0)}_{i}{}^a(x)+u^ba^{(1)}_{i}{}^a{}_b(x)+\tfrac12u^bu^ca^{(2)}_{i}{}^a{}_{bc}(x)+\dots
\eeq
and using \eqref{dela}, we can read off the infinitesimal transformations of $a^{(0)}_{i}{}^a$ and $a^{(1)}_{i}{}^a{}_b$ under the embedding symmetry
\beqn\label{deltaazero}
\delta_{\un\xi}a^{(0)}_{i}{}^a(x)&=&
\xi_{(0)}^j\pa_j a^{(0)}_{i}{}^a
+a^{(0)}_{j}{}^a\pa_i\xi_{(0)}^j
-\xi_{(1)}{}^a{}_ba^{(0)}_{i}{}^b
-\pa_i\xi_{(0)}^a
+a^{(1)}_{i}{}^a{}_b\xi_{(0)}^b,
\\
\label{deltaaone}
\delta_{\un\xi} a^{(1)}_{i}{}^a{}_b(x)&=& 
\xi_{(0)}^j\pa_ja^{(1)}_{i}{}^a{}_b
+a^{(1)}_{j}{}^a{}_b\pa_i\xi_{(0)}^j
-\pa_i\xi_{(1)}{}^a{}_b
+ a^{(1)}_{i}{}^a{}_c\xi_{(1)}{}^c{}_b
-\xi_{(1)}{}^a{}_ca^{(1)}_{i}{}^c{}_b
+a^{(2)}_{i}{}^a{}_{bc}\xi^c_{(0)}.
\eeqn
All of the $a_i^{(k)}$ for $k\geq 2$ transform linearly under ${\cal A}_k$, and thus it is consistent to set these to zero.\footnote{Note that the translation subalgebra $\RR^k$ couples  $a^{(k-1)}_i{}^a$ to $a^{(k)}_i{}^a$ because translations can be interpreted as moving $S$ in the normal direction. Fixing to the embedding $\phi_0$, and setting $\xi^a_{(0)}$ and $a_i^{(0)}{}^a$ to zero then cleanly leaves  \eqref{deltaaone} as just the transformation of a $GL(k,\RR)$ connection.}
In each of these equations the first two terms are present because $a^{(0)}_{i}{}^a(x)$ and $a^{(1)}_{i}{}^a{}_b(x)$ transform covariantly under $Diff(S)$. In \eqref{deltaazero}, the third term is present because $a^{(0)}_{i}{}^a(x)$ transforms as a vector under $GL(k,\RR)$. The last two terms in \eqref{deltaazero} and the last three terms in \eqref{deltaaone} are $GL(k,\RR)$-covariant derivatives, indicating that $a^{(0)}_{i}{}^a(x)$ is a connection for $\RR^k$ and $a^{(1)}_{i}{}^a{}_b(x)$ a connection for $GL(k,\RR)$. As such, it is convenient to introduce notation for the $GL(k,\RR)$-covariant derivatives,
\beqn
{\cal D}_i\xi_{(0)}^a&\equiv&\pa_i\xi_{(0)}^a-a^{(1)}_{i}{}^a{}_b\xi_{(0)}^b,
\\
{\cal D}_i\xi_{(1)}{}^a{}_b&\equiv&\pa_i\xi_{(1)}{}^a{}_b
-a^{(1)}_{i}{}^a{}_c\xi_{(1)}{}^c{}_b
+\xi_{(1)}{}^a{}_ca^{(1)}_{i}{}^c{}_b
=\pa_i\xi_{(1)}{}^a{}_b-\big[a^{(1)}_{i},\xi_{(1)}\big]^a{}_b.
\eeqn
This suggests that the embedding symmetries, linearly realized in $M$, can be reformulated intrinsically, without any reference to $M$ itself. In this sense, this is the realization of a holographic nature of gauge theories. In this section, we showed  the geometric and then the algebraic aspects of a gravitational theory in the presence of an isolated embedded surface $S$. Note though that the discussion has so far been entirely off-shell, without even referring to a metric on $M$. In the next section, we will introduce a metric and connection adapted to the embedding.

\section{Adapted Metrics}\label{sec:metric}

In the last section, we have shown that $a_i^a$ transforms as a connection under the maximal embedding algebra of diffeomorphisms ${\cal A}_k$. At this stage, the reader may be puzzled by this result for several reasons. First, connections usually show up in gravitational theories as derivatives of a metric, and so far we have not made reference to a metric at all. One should in fact regard this as the power of the method. Another puzzle concerns how the $GL(k,\RR)$ connection may show up given that the forms $n^a$ are defined to pull back to zero on $S$, and that would seem to make the connection disappear. In this section, we will explain carefully how these puzzles are resolved.  While in the previous section we focused on the embedding, in this section we start from the bulk endowed with an arbitrary metric and adapt its description to the embedding. 

Given that we have the forms $n^a$ in hand, without loss of generality we can introduce a metric for $M$ in the adapted coordinates as
\beq\label{randers}
g=h_{ab}(u,x)n^a\otimes n^b+\gamma_{ij}(u,x)dx^i\otimes dx^j.
\eeq
In this line element, since the $n^a$ pull back to zero on $S$, we see that $\gamma_{ij}$  pulls back to a metric on $S$, while $h_{ab}(u,x)$ can be thought of as a metric on the normal fibre at any given point on $S$.

The metric \eqref{randers}, especially in the codimension-1 case, is familiar; in some contexts it would be referred to as of the Randers-Papapetrou parameterization, \cite{PhysRev.59.195, Papapetrou:1966zz, Gibbons:2008zi}. 
In this form of the metric, particular emphasis is given to the spatial forms $dx^i$. In this regard, eq. \eqref{randers} can be viewed as a specific form, naturally restricted to the problem at hand, of the standard  Newman-Penrose formalism \cite{Newman:1961qr}, and in particular the subsequent Geroch-Held-Penrose construction \cite{Geroch:1973am}, as arising for instance in the treatment of asymptotic properties of a spacetime (see  \cite{Adamo:2009vu} and references therein).

Note that this form of the metric differs from the form usually used in the context of foliations (such as in the ADM formalism \cite{Arnowitt:1962hi}), 
\beq\label{zermelo}
g=\tilde h_{ab} du^a\otimes du^b+\tilde\gamma_{ij}(dx^i+N^i_adu^a)\otimes (dx^j+N^j_b du^b).
\eeq
In some contexts, this is referred to as a Zermelo form of the metric \cite{https://doi.org/10.1002/zamm.19310110205}.
At any generic point in $M$, the two metrics are equivalent and simply correspond to a reshuffling of notation. However, \eqref{zermelo} is a good parameterization in the context of a foliated manifold, because the constituents transform in a compact way under $u'=u'(u), x'=x'(u,x)$ which correspond to arbitrary and independent diffeomorphisms on each leaf of the foliation along with reparameterizations of the leaf labels. On the other hand, the metric \eqref{randers} is preferred in the context of an isolated embedded surface because as we will now show, the (pull-backs of the) constituents transform as intrinsically defined tensors or connections on the surface with respect to the diffeomorphisms $u'=u'(u,x), x'=x'(x)$. We interpret such diffeomorphisms to correspond to the diffeomorphisms of the surface, along with local reparameterizations of $u^a$. Furthermore, the diffeomorphisms generating the maximal embedding algebra ${\cal A}_k$ are a subset of this form, and not of the form natural to the foliation. It is the presence of the surface itself that distinguishes the two forms of the metric; we have in fact already seen the first manifestation of this, that the translation $\RR^k$ symmetry is spontaneously broken. In the recent literature on corners \cite{Donnelly:2020xgu}, a metric of the form \eqref{zermelo} was chosen. The data forming the connection for the corner symmetry are then to be found within the various bulk Christoffel symbols. Although this leads eventually to correct results, the foliation setup blurs the geometric organization of the problem. As we will now uncover, choosing \eqref{randers} and working with an isolated embedding bypasses this problem yielding the sought-for connection directly in the line element, with a clear geometric interpretation as the connection defining the adapted split of the tangent bundle.

To explore this further, let us first note that the coordinate components are
\beq
g_{ab}=h_{ab},\qquad g_{aj}=-h_{ab}a_j^b,\qquad g_{ij}=\gamma_{ij}+h_{ab}a_i^aa_j^b,
\eeq
and the inverse is
\beq
g^{ab}=h^{ab}+\gamma^{ij}a_i^aa_j^b,\qquad g^{aj}=\gamma^{jk}a_k^a,\qquad g^{ij}=\gamma^{ij},
\eeq
where $h^{ab}h_{bc}=\delta^a{}_c$ and $\gamma^{ij}\gamma_{jk}=\delta^i{}_k$.

Infinitesimal diffeomorphisms act as usual as
\beqn
({\cal L}_{\un\xi}g)_{ab}&=&
\xi^c\pa_c h_{ab}+\xi^i\pa_i h_{ab}+h_{bc}B^c{}_a+h_{ac}B^c{}_b
\\
({\cal L}_{\un\xi}g)_{ai}
&=&
-\xi^c\pa_c (h_{ab}a_i^b)
-h_{cb}a_i^b\pa_a\xi^c
+h_{ac}\pa_i\xi^c
-\xi^j\pa_j (h_{ab}a_i^b)
+(\gamma_{ij}+h_{bc}a_i^ba_j^c)\pa_a\xi^j
-h_{ab}a_j^b\pa_i\xi^j
\\
({\cal L}_{\un\xi}g)_{ij}
&=&
\xi^a\pa_a \gamma_{ij}
+\xi^a\pa_a (h_{bc}a_i^ba_j^c)
-h_{ab}a_j^b\pa_i\xi^a
-h_{ab}a_i^b\pa_j\xi^a
+\xi^k\pa_k(h_{bc}a_i^ba_j^c)
\nonumber\\&&
+h_{bc}a_k^ba_j^c\pa_i\xi^k
+h_{bc}a_k^ba_i^c\pa_j\xi^k
+\xi^k\pa_k \gamma_{ij}
+\gamma_{k j}\pa_i\xi^k
+\gamma_{ki}\pa_j\xi^k,
\eeqn
where for compactness we introduced
\beq
B^a{}_b\equiv \pa_b\xi^a-a_j^a\pa_b\xi^j.
\eeq
These bulk Lie derivatives induce the following changes in the metric consitituents
\beqn
\delta_{\un\xi}h_{ab}&=&
\xi^j\pa_j h_{ab}+\xi^c\pa_c h_{ab}+h_{bc}B^c{}_a+h_{ac}B^c{}_b
\\
\delta_{\un\xi}a_i^b&=&
\xi^j\pa_ja_i^b+a_j^b\pa_i\xi^j
-B^b{}_ca_i^c-\pa_i\xi^b+\xi^c\pa_c a_i^b
-h^{ba}\gamma_{ij}\pa_a\xi^j
\\
\delta_{\un\xi}\gamma_{ij}
&=&
\xi^k\pa_k \gamma_{ij}
+\xi^a\pa_a \gamma_{ij}
+\gamma_{k j}D_i\xi^k
+\gamma_{ki}D_j\xi^k.
\eeqn
We now restrict our attention to the maximal embedding algebra ${\cal A}_k$,
\beq\label{surfpresdiff2}
\xi^j=\xi_{(0)}^j(x),\qquad
\xi^a=\xi_{(0)}^a(x)+u^b\xi_{(1)}{}^a{}_b(x),
\eeq
and expand the metric constituents around $u^a=0$. An important question arises at this point, concerning the behavior in $u^a$ of the metric constituents as we approach $u^a=0$. In considering embedding at (conformally) infinite distance in the bulk, it is well-known that there is a coordinate-independent pole structure, so that the induced metric structure is actually an induced conformal class \cite{Penrose:1962ij, Penrose:2011wy}. The coordinate realization of this phenomenon is that the metric has a pole in terms of the extrinsic coordinates (as e.g., for $k=1$ in AdS, the holographic coordinate). Although asymptotic surfaces are of significant importance in many contexts, and we plan to return to them in future work, we will in the rest of the current paper consider embedded surfaces that are at finite proper distance, imposing therefore that the metric constituents expand smoothly around $u^a=0$. This has deep consequences for the Noether charges discussed in the next section, especially in the relationship to the generalization of the Weyl symmetry and associated charges \cite{Ciambelli:2019bzz, Alessio:2020ioh}. Thus, working in this context and restricting our attention to \eqref{surfpresdiff2}, we have
\beqn
\delta_{\un{\xi}}h^{(0)}_{ab}&=&
\xi_{(0)}^j\pa_j h^{(0)}_{ab}
+\Big(h^{(0)}_{bc}\xi_{(1)}{}^c{}_a+h^{(0)}_{ac}\xi_{(1)}{}^c{}_b\Big)
+\xi_{(0)}^ch^{(1)}_{ab,c}\label{h0}
\\
\delta_{\un{\xi}}a_i^{(0)}{}^b&=&
\Big(\xi_{(0)}^j\pa_ja_i^{(0)}{}^b
+a_j^{(0)}{}^b\pa_i\xi_{(0)}^j\Big)
-\xi_{(1)}{}^b{}_ca_i^{(0)}{}^c
+\Big(-\pa_i\xi_{(0)}^b+\xi_{(0)}^c a_i^{(1)}{}^b{}_c\Big)\label{a0}
\\
\delta_{\un{\xi}}a_i^{(1)}{}^a{}_b&=&
\Big(\xi_{(0)}^j\pa_ja_i^{(1)}{}^a{}_b
+a_j^{(1)}{}^a{}_b\pa_i\xi_{(0)}^j\Big)
+\Big(-\pa_i\xi_{(1)}{}^a{}_b
+a_i^{(1)}{}^a{}_c\xi_{(1)}{}^c{}_b
-\xi_{(1)}{}^a{}_ca_i^{(1)}{}^c{}_b
\Big)
+\xi_{(0)}^c a_i^{(2)}{}^a{}_{bc}\label{a1}
\\
\delta_{\un{\xi}}\gamma^{(0)}_{ij}
&=&
\Big(\xi_{(0)}^k\pa_k \gamma^{(0)}_{ij}
+\gamma^{(0)}_{k j}\pa_i\xi_{(0)}^k
+\gamma^{(0)}_{ki}\pa_j\xi_{(0)}^k\Big)
+\xi_{(0)}^c \gamma^{(1)}_{ij,c},\label{g0}
\eeqn
and so on. 
We have grouped together terms that correspond to the transformations under the algebras
 $\dS$, $\mathfrak{gl}(k,\RR)$ and $\RR^k$ respectively. 
So we see that under the restricted diffeomorphisms, the metric constituents transform appropriately under the algebra ${\cal A}_k$ of the group $(Diff(S)\ltimes GL(k,\RR))\ltimes \RR^k$: $h^{(0)}_{ab}$ and $\gamma^{(0)}_{ij}$ are respectively scalars and a tensor with respect to $Diff(S)$ and a tensor and scalars under $GL(k,\RR)$, with translations coupling them to their next order in $u^a$. Remarkably, and expectedly, \eqref{a0} and \eqref{a1} are the same as \eqref{deltaazero} and \eqref{deltaaone}, showing as advertised that the connection for the maximal symmetry is encoded in constituents of the bulk  metric in the parameterization \eqref{randers}. Strictly speaking, the connections we found here are only the ones for the $GL(k,\RR)$ and $\RR^k$ symmetries. The connection for $Diff(S)$ arises as an induced connection from the bulk Levi-Civita connection. The details of this are discussed in Appendix \ref{AppB}. We observe that \eqref{a0} can be rephrased from the point of view of forms pulled back to $S$ and is then consistent with our discussion in Section \ref{sec2.1}. A translation of the normal directions changes the embedding. Indeed, starting with the embedding  $\phi_0$, and hence $a^a_\alpha(\sigma)=0$, and performing a translation, the new embedding corresponds to $u^a(\sigma)=-\xi^a_{(0)}(x(\sigma))$, and the translation connection correctly becomes $a^a_\alpha(\sigma)=-\pa_\alpha\xi^a_{(0)}(x(\sigma))$.

The adapted split of the tangent bundle $TM$ discussed in Section \ref{sec2.1} introduces a non-coordinate basis in $TM$ such that the rotation coefficients encode the relevant geometric quantities of the problem. We have
\beqn
\big[\un D_i,\un D_j\big]&=& f_{ij}^a(u,x)\un\pa_a,\\
\big[ \un D_i,\un\pa_a\big]&=& -\pa_aa^b_i(u,x)\un\pa_b,
\eeqn
with $f_{ij}^a(u,x)$ given in eq. \eqref{curv}. Borrowing vocabulary proper to hydrodynamics, we might refer to $f_{ij}^a(u,x)$ as the vorticity (related to integrability of the horizontal subbundle $H$) and $\varphi_i{}^b{}_a=-\pa_aa^b_i(u,x)$ the acceleration of the vertical congruence $\un\pa_a$. Expanding near $u^a=0$ we obtain
\beqn
\big[\un D_i,\un D_j\big]&=& W^a{}_b{}_{ij}(x)\ u^b\un\pa_a +...,\\
\big[ \un D_i,\un\pa_a\big]&=&-a^{(1)}_i{}^b{}_a(x)\ \un\pa_b+...,
\eeqn
where we have defined the ${\cal A}_k$-tensor
\beq
W^a{}_b{}_{ij}=D_ia^{(1)}_j{}^a{}_b-D_ja^{(1)}_i{}^a{}_b.
\eeq
So the leading order of the acceleration  and the vorticity are the $GL(k,\RR)$ connection and its field strength. The latter plays a pivotal role in the analysis performed in \cite{Donnelly:2020xgu}. Here, we appreciate its geometrical significance as the leading order in the obstruction to the integrability of the horizontal sub-bundle $H$ of $TM$. The geometric content of the bulk established, we now proceed by specializing to $k=2$, where the maximal embedding algebra is further realized in terms of Noether charges.

\section{Corner Charges and Algebra}\label{sec:dynamics}

In this section we focus on the case $k=2$, where the $n=d-2$ dimensional embedded surfaces are corners. It is in this context that we can make the link between the vector fields studied above and the gauge charges of a particular dynamical theory. We will confine our attention here to the Noether charges of Einstein-Hilbert theory in the bulk, although other dynamics, and other charges, would be also fruitful to study. There are two main results. First, we show that the Noether charges realize on phase space the generators of the full maximal embedding algebra; however, at least for the case of finite proper distance isolated corners that we consider here explicitly, only $\mathfrak{sl}(2,\RR)\subset \mathfrak{gl}(2,\RR)$ is realized in the Einstein-Hilbert theory. Second, we show that the Noether charges give a representation of the full maximal embedding algebra, normal translations included, without central extension. To realize this result, it is necessary to carefully treat the embedding of the corners. We take these results to indicate that the Noether charge should be interpreted as a corner charge associated with the maximal embedding algebra. We then conclude this section by showing how the general construction applies to the specific example of near-horizon symmetries.

\subsection{Noether Charges}

We specialize to the Einstein-Hilbert theory and consider the corresponding Noether diffeomorphism charges associated to vector fields generating $\dM$ in the bulk.
It is well-known that these have support in codimension-$2$ and in our context it is natural to consider their expression as an integral over an isolated embedded corner $\phi:S\to M$. 
For a vector field $\un\xi$ on $M$, the corresponding Noether charge is given by
\beq\label{firstCH}
H_{\un\xi}=\int_S \phi^*(*d g(\un\xi,\cdot)),
\eeq
where $\phi^*$ is the pullback of the embedding $\phi$, $*$ is the bulk Hodge dual and $d$ the exterior derivative of the one form $g(\un\xi,\cdot)=i_{\un\xi}g$. We emphasize that the embedding is part of the definition of the charge. We will regard vector fields as field-independent when their components in a coordinate basis are field-independent.

Suppose we have local coordinates $y^M$ on $M$ and a metric $g=g_{MN}(y)dy^M\otimes dy^N$. In these coordinates we write a field independent infinitesimal vector field $\un\xi=\xi^M(y)\un\pa_M$ and so we compute
\beqn
g(\un\xi,\cdot)&=&
g_{MN}(y)\xi^M(y)dy^N
\\
dg(\un\xi,\cdot)&=&
\frac12\Big(\pa_P(g_{NM}\xi^M(y))-\pa_N(g_{PM}\xi^M(y))\Big)dy^P\wedge dy^N \ \ \equiv \ \ \frac12 \kappa_{PN}(y)dy^P\wedge dy^N.
\eeqn
We then define the bulk $n$-form (with the Levi-Civita symbol $\varepsilon$)
\beqn
K[\un\xi,g,y]\equiv
*dg(\un\xi,\cdot)
&=& \frac12\sqrt{-\det g(y)}g^{M_1P}(y)\kappa_{PN}(y)g^{NM_2}(y)\frac{1}{n!}\varepsilon_{M_1M_2M_3...M_d}dy^{M_3}\wedge ... \wedge dy^{M_d}
\\
&\equiv& \frac12k_{\un\xi}^{M_1M_2}(y)\frac{1}{n!}\varepsilon_{M_1M_2M_3...M_d}dy^{M_3}\wedge ... \wedge dy^{M_d}.
\eeqn
While we work in fixed bulk coordinates $y^M$,  in these coordinates $S$ is described by the embedding $y^M=y^M(\sigma)$, where $\sigma^\alpha$ are coordinates on $S$. In order to integrate this $n$-form, we must first pull it back to $S$. We then have
\beqn
\phi^*(K[\un\xi,g,y])
=  \frac12k_{\un\xi}^{M_1M_2}(y(\sigma))\ (\pa_{\alpha_1} y^{M_3}(\sigma))...(\pa_{\alpha_n} y^{M_d}(\sigma))\frac{1}{n!}\varepsilon^{\alpha_1...\alpha_n}\varepsilon_{M_1M_2M_3...M_d}vol_0\label{genEmb},
\eeqn
where for brevity we have written
$vol_0=\frac{1}{n!}\varepsilon_{\alpha_1...\alpha_n}d\sigma^{\alpha_1}\wedge...\wedge d\sigma^{\alpha_n}$.

Now let us rewrite this expression in the adapted coordinates $y^M=(u^a,x^i)$. We find
\beqn
\phi^*(K[\un\xi,g,y])
&=&  \varepsilon_{ab}\Big[\frac12k_{\un\xi}^{ab}(y(\sigma))\ \pa_{\alpha_1} x^{i_1}(\sigma)\pa_{\alpha_2} x^{i_2}(\sigma)
- n k_{\un\xi}^{ai_1}(y(\sigma))\ \pa_{\alpha_1} u^{b}(\sigma)\pa_{\alpha_2} x^{i_2}(\sigma)
\\
&&+  \frac12\frac{n(n-1)}{2}k_{\un\xi}^{i_1i_2}(y(\sigma))\ \pa_{\alpha_1} u^{a}(\sigma)\pa_{\alpha_2} u^{b}(\sigma)\Big]\frac{1}{n!}\varepsilon^{\alpha_1...\alpha_n}(\pa_{\alpha_3} x^{i_3}(\sigma)...\pa_{\alpha_n} x^{i_n}(\sigma))\varepsilon_{i_1...i_n}vol_0.
\nonumber\eeqn
This can be simplified further by recalling that the embedding satisfies
$
\pa_\alpha u^a(\sigma)=\pa_\alpha x^i(\sigma) a^a_i(u(\sigma),x(\sigma))
$. After some algebra, we find 
\beqn
\phi^*(K[\un\xi,g,y])
&=&  \frac12\sqrt{-\det h}\ \varepsilon_{ab}
h^{ac}\kappa_{cd}h^{db}
{\cal J}\sqrt{\det\gamma}vol_0
\eeqn
where
\beqn
\label{formoff1}
\kappa_{cd}
=\pa_c(h_{de}\xi_V^e)-\pa_d(h_{ce}\xi_V^e)
\eeqn
and
${\cal J}=\frac{1}{n!}\varepsilon^{\alpha_1...\alpha_n}\pa_{\alpha_1} x^{i_1}(\sigma)...\pa_{\alpha_n} x^{i_n}(\sigma)\varepsilon_{i_1...i_n}$.
Noting that ${\cal J}\sqrt{\det\gamma}\ vol_0=vol_S$, we finally obtain\footnote{Note added: In a subsequent paper \cite{Freidel:2021cbc}, expressions for the bulk forms entering the Noether charges have been obtained in the Zermelo parameterization, in a trivial embedding. } 
\beqn
\phi^*(K[\un\xi,g,y])
=  \sqrt{-\det h}\ h^{ca}\varepsilon_{ab}h^{bd}
\pa_c(h_{de}\xi_V^e)\
vol_S,
\eeqn
where all quantities are to be evaluated at $y(\sigma)$. For the sake of brevity, we introduced here the notation $\xi_V^e\equiv\xi^e-a^e_j\xi^j$ (which in fact correspond to the components of a vertical vector field).

It is instructive to evaluate the charges in the trivial embedding $\phi_0$, given by $y_0^M(\sigma)=(0,\delta^i_\alpha\sigma^\alpha)$. 
One finds that the charge can be simply expressed by making use of the notation introduced in eq. \eqref{exp}, and we  find 
\beqn
H_{\un\xi}
&=&\int_S vol_S\ \sqrt{-\det h^{(0)}}\ h_{(0)}^{ae}\varepsilon_{ec}\Big(\xi_{(1)}{}^c{}_a-a^{(1)}_j{}^c{}_a\xi_{(0)}^j+h_{(0)}^{cb}\xi_{(0)}^dh^{(1)}_{db,a}\Big)\label{zeropullbackres}
\\
&\equiv&\int_S vol_S\ \Big(\xi_{(1)}{}^a{}_bN^b{}_a+\xi_{(0)}^jb_j+\xi_{(0)}^ap_a\Big)\label{fullcharge},
\eeqn
(all the functions appearing are functions on the corner, i.e., of $\sigma^\alpha$)
where we have introduced
\beqn\label{N}
N^b{}_a &=& \sqrt{-\det h^{(0)}}\ h_{(0)}^{bc}\varepsilon_{ca}\\\label{b}
b_j &=& -N^b{}_aa^{(1)}_j{}^a{}_b\\
p_d &=& \tfrac12 N^{a}{}_{c}h_{(0)}^{cb}(h^{(1)}_{db,a}-h^{(1)}_{da,b}).\label{d}
\eeqn
The first important remark is that only $\xi_{(1)}{}^a{}_b$, $\xi^j_{(0)}$ and $\xi^a_{(0)}$ contribute in \eqref{fullcharge}, showing that only the vector fields of the maximal embedding algebra for $k=2$ (that is, ${\cal A}_2$) give rise to non-vanishing Noether charges. This is consistent with the fact established above that these vector fields generate the maximal closed sub-algebra of $\dM$, with all the rest of $\dM$ acting trivially at the corner.
$N^a{}_b$ is clearly associated with $\mathfrak{gl}(2,\RR)$, but we notice that it is a traceless ${\cal A}_2$ tensor so only the $\mathfrak{sl}(2,\RR)\subset \mathfrak{gl}(2,\RR)$ is realized on the Einstein-Hilbert phase space, as the charge algebra computation will confirm shortly. This is our aforementioned result: the Noether charges of the Einstein-Hilbert theory for a manifold $M$ with finite proper distance isolated corner $S$ realizes only the  $\tilde{\cal A}_2=\Big(\dS\loplus \mathfrak{sl}(2,\RR)\Big)\loplus \RR^2$ subalgebra of ${\cal A}_2$. We stress that this result concerns only finite proper distance corners. The diffeomorphisms on $S$ are generated on phase space by $b_j$ while the normal translations are generated by $p_d$.
The first two terms in \eqref{fullcharge} were recently derived also in Ref. \cite{Donnelly:2020xgu}, and play an important role in the coadjoint representation\footnote{The coadjoint orbits of $\mathfrak{bms}_3$ and $\mathfrak{bms}_4$ have been studied in Refs. \cite{Barnich:2015uva,Barnich:2021dta}.} of $\dS$ and $\mathfrak{sl}(2,\RR)$. The normal translations were not considered in that work, but their effects will be crucial in the Noether charge algebra.

\subsection{Charge Algebra}

To evaluate the algebra, we will examine $\delta_{\un\eta}H_{\un\xi}$. This is subtle in the general case because of the normal translations: we must consider variations of the bulk fields, but also we must consider a corresponding variation in the embedding of the surface that is part of the definition of the charge. The latter is in keeping with the idea that the embedding gives rise to new degrees of freedom \cite{Donnelly:2016auv,Freidel:2019ofr} that otherwise would have been pure gauge in the absence of the corner. To compute the charge algebra, one should evaluate the variation of the fields in $K$ holding fixed the parameters $\un\xi$ and $\un\eta$, if the latter are field independent, i.e., they do not depend on the metric constituents. This is the case if one does not specify a gauge for the bulk metric and does not impose specific boundary conditions. Indeed, requiring the vector fields to preserve a gauge or specific boundary conditions imposes constraints on them that can result in residual vector fields that depend explicitly on the metric. In the derivation of the charge algebra below, to avoid confusion, we will at first suppose that $\un\xi$ is field independent. After deriving the algebra, we will show that promoting the vectors to be field dependent has a straightforward impact. Therefore,  we have 
\beq\label{LieDE}
\delta_{\un\eta}K[\un\xi,g,y]=K[\un\xi,\delta_{\un\eta}g,y]=K[\un\xi,{\cal L}_{\un\eta}g,y].
\eeq
If the transformations that we are considering did not change the embedding, then we would just have 
\beq\label{ChAl}
\delta_{\un\eta}H_{\un\xi}=\delta_{\un\eta}\int_S\phi^*(K[\un\xi,g,y])=\int_S\phi^*(\delta_{\un\eta}K[\un\xi,g,y])=\int_S\phi^*(K[\un\xi,{\cal L}_{\un\eta}g,y]).
\eeq
Here, however, we are interested in transformations that change the embedding as well, so we are not allowed to commute $\delta_{\un\eta}$ past $\phi^*$. In general we then have an extra contribution that we write as
\beq
\delta_{\un\eta}H_{\un\xi}
=\int_S\phi^*(\delta_{\un\eta}K[\un\xi,g,y])+\int_S(\delta_{\un\eta}\phi^*)(K[\un\xi,g,y]).
\eeq
Treating the bulk diffeomorphism passively, the additional effect $\delta_{\un\eta}\phi^*$ comes from the fact that the surface $S$ is located at $y'{}^M=y'{}^M(\sigma)$ in the $y'$ coordinates, while in the $y$-coordinates it is described by $y^M=y^M(\sigma)$. We remark that the charges, as well as the algebra, are defined for any choice of embedding. However, since local translations are included in the set of generators and, as already discussed, they change the initial choice of embedding, it is crucial to vary it in this computation. In this sense, $\phi$ is on a similar footing to the metric $g$. Infinitesimally, for $y'{}^M(y)\simeq y^M-\eta^M(y)$, this reads\footnote{Since this result is key to our derivation of the algebra, we show it explicitly in appendix \ref{AppA}.}
\beq\label{delphi}
\int_S(\delta_{\un\eta}\phi^*)(K[\un\xi,g,y])=\int_S\phi^*\Big(K[\un\xi,g,y-\eta]-K[\un\xi,g,y]\Big).
\eeq
Gathering the various results, we arrive at the final expression
\beq\label{finCH}
\delta_{\un\eta}H_{\un\xi}=\int_S\phi^*\Big(K[\un\xi,{\cal L}_{\un\eta}g,y])+K[\un\xi,g,y-\eta]-K[\un\xi,g,y]\Big),
\eeq
valid for a generic embedding and an arbitrary vector field $\un\xi$. This expression is useful in practice because the passive interpretation adopted here allows us to compute the effects of the change of embedding directly in the bulk and then utilize only the unmodified embedding.\footnote{We also observe that this result, generalizing accordingly the $n$-form $K$, can be applied to higher form symmetries, where the codimension can be higher than two.} 

We finally have all the ingredients to obtain $\delta_{\un\eta}H_{\un\xi}$. We will compute this in the embedding $\phi_0$ corresponding to $y^M=y_0^M(\sigma)=(0,\delta^i_\alpha\sigma^\alpha)$. First, since $K[\un\xi,g,y]$ is a bulk $n$-form that is a specific functional of the metric $g$, it is an invariant quantity under arbitrary bulk diffeomorphisms
\beq
K[\un\xi',g',y']=K[\un\xi,g,y],
\eeq
where $\un\xi'$ and $g'$ are transformed  generators and  fields, while $y'$ are the new coordinates. Using $y'{}^M(y)\simeq y^M-\eta^M(y)$, the variation at the same point is as usual the Lie derivative, that is,
\beqn
K[\un\xi',g',y]-K[\un\xi,g,y]
\simeq {\cal L}_{\un\eta}K[\un\xi,g,y]
=
K[\un\xi,{\cal L}_{\un\eta}g,y]+K[{\cal L}_{\un\eta}\un\xi,g,y].
\eeqn
An explicit calculation then gives
\beqn
{\cal L}_{\un\eta}K[\un\xi,g,y]&=&
di_{\un\eta}K[\un\xi,g,y]+i_{\un\eta}dK[\un\xi,g,y]\\
&=&
\frac12 k_{\un\xi}^{PQ}(\pa_R\eta^S) \frac{1}{(n-1)!} \varepsilon_{PQSM_2...M_n}dy^R\wedge dy^{M_2}\wedge...\wedge dy^{M_n}\\
&&+\frac12 \eta^R(\pa_Rk_{\un\xi}^{PQ})\frac{1}{n!} \varepsilon_{PQM_1...M_n} dy^{M_1}\wedge...\wedge dy^{M_n}\nonumber.
\eeqn
We now pull this back using $\phi^*_0$. After some simple manipulations we arrive at
\beqn
\phi_0^*({\cal L}_{\un\eta}K[\un\xi,g,y])&=&
\frac12 \delta_{\alpha_1}^{i_1}k_{\un\xi}^{PQ}(\pa_{i_1}\eta^S) \frac{1}{(n-1)!} \varepsilon_{PQS\alpha_2...\alpha_n} d\sigma^{\alpha_1}\wedge ...\wedge d\sigma^{\alpha_n}
+\frac12 \eta^R\pa_R(\varepsilon_{ab}k_{\un\xi}^{ab}) vol_0\\
&=&
\Big[
\varepsilon_{ab}k_{\un\xi}^{ja}\pa_{j}\eta^b 
+\tfrac12 \eta^c\pa_c(\varepsilon_{ab}k_{\un\xi}^{ab})
+\tfrac12 \pa_{j}\Big(\eta^j\varepsilon_{ab}k_{\un\xi}^{ab}\Big)
\Big](y_{0}(\sigma))\ vol_0\label{uhohextraterms}.
\eeqn
So the first term of the integrand on the right hand side of \eqref{finCH} for the embedding $\phi^*_0$ is
\beq
\phi^*_0(K[\un\xi,{\cal L}_{\un\eta}g,y])=\Big[
\varepsilon_{ab}k_{\un\xi}^{ja}\pa_{j}\eta^b 
+\tfrac12 \eta^c\pa_c(\varepsilon_{ab}k_{\un\xi}^{ab})
+\tfrac12 \pa_{j}\Big(\eta^j\varepsilon_{ab}k_{\un\xi}^{ab}\Big)
\Big](y_{0}(\sigma))\ vol_0-\phi^*_0(K[{\cal L}_{\un\eta}\un\xi,g,y]).
\eeq
The last term in this expression is nothing but the integrand of the Noether charge associated to the Lie bracket of the vector fields
\beqn
-\int_S \phi_0^*(K[{\cal L}_{\un\eta}\un\xi,g,y])= H_{[\un\xi,\un\eta]}.
\eeqn
We now use our result \eqref{fullcharge} for $H_{[\un\xi,\un\eta]}$, where $[\un\xi,\un\eta]_{(0)}^j$, $[\un\xi,\un\eta]_{(0)}^a$ and $[\un\xi,\un\eta]_{(1)}{}^a{}_b$ are the expressions given in  \eqref{LieMsub}, reported explicitly below to clarify notation
\beqn
\big[\un\xi,\un\eta\big]_{(0)}^j&=&\big[\un{\hat\xi}_{(0)},\un{\hat\eta}_{(0)}\big]^j
\label{Scomm}\\
\big[\un\xi,\un\eta\big]_{(0)}^a&=&
\un{\hat\xi}_{(0)}(\eta_{(0)}^a)
-\un{\hat\eta}_{(0)}(\xi_{(0)}^a)
-\xi_{(1)}{}^a{}_b\eta_{(0)}^b
+\eta_{(1)}{}^a{}_b\xi_{(0)}^b
\\
\big[\un\xi,\un\eta\big]_{(1)}{}^a{}_b&=&
-\big[\xi_{(1)},\eta_{(1)}\big]^a{}_b
+\un{\hat\xi}_{(0)}(\eta_{(1)}{}^a{}_b)
-\un{\hat\eta}_{(0)}(\xi_{(1)}{}^a{}_b).
\label{Gcomm}
\eeqn

We then calculate the remaining terms in \eqref{finCH}, i.e., $\phi^*(K[\un\xi,g,y-\eta])-\phi^*(K[\un\xi,g,y])$, for the embedding $\phi^*_0$, which yields
\beqn\label{RHS}
\phi_0^*(K[\un\xi,g,y-\eta]-K[\un\xi,g,y])
=-\Big[
\varepsilon_{ab}k_{\un\xi}^{ja}\pa_{j}\eta^b 
+\frac12 \eta^c\pa_c(\varepsilon_{ab}k_{\un\xi}^{ab})
+\frac12 \pa_{j}\Big(\eta^j\varepsilon_{ab}k_{\un\xi}^{ab}\Big)
\Big](y_{0}(\sigma))\ vol_0.
\eeqn
This contribution cancels exactly \eqref{uhohextraterms}, leading to the final result
\newcommand{\brac}[2]{\{\kern-2pt[ #1,#2]\kern-2pt\}}
\beq
\delta_{\un\eta}H_{\un\xi}=H_{[\un\xi,\un\eta]}
=\brac{ H_{\un\xi}}{H_{\un\eta}},
\eeq
where we have introduced a bracket notation.
Defining the individual charges by projecting the vector field onto individual components,
\beq
N_{\un\xi}=\int_S vol_S\ \xi_{(1)}{}^a{}_b(\sigma)N^b{}_a,\qquad
b_{\un\xi}=\int_S vol_S\ \xi_{(0)}^j(\sigma)b_j,\qquad
p_{\un\xi}=\int_S vol_S\ \xi_{(0)}^a(\sigma)p_a,
\eeq
we then read off the brackets of these charges
\beqn
\brac{b_{\un\xi}}{b_{\un\eta}}=b_{[\un\xi,\un\eta]}=b_{[\hat{\un\xi},\hat{\un\eta}]},
&\brac{b_{\un\xi}}{N_{\un\eta}}=N_{[\un\xi,\un\eta]}=N_{\hat{\un\xi}_{(0)}(\eta_{(1)})},
&\brac{b_{\un\xi}}{p_{\un\eta}}=p_{[\un\xi,\un\eta]}=p_{\hat{\un\xi}_{(0)}(\eta_{(0)})}
\\ 
&\brac{N_{\un\xi}}{N_{\un\eta}}=N_{[\un\xi,\un\eta]}=-N_{[\xi_{(1)},\eta_{(1)}]}, 
&\brac{N_{\un\xi}}{p_{\un\eta}}=p_{[\un\xi,\un\eta]}=-p_{\xi_{(1)}\cdot\eta_{(0)}}
\\
&&
\brac{p_{\un\xi}}{p_{\un\eta}}=0.
\eeqn
where the brackets in the middle expressions have been evaluated by consulting eqs. (\ref{Scomm}--\ref{Gcomm}).
This structure is of course consistent with the identification of the group in terms of semi-direct products, showing that the brackets of the Noether charges give a representation of the algebra $\tilde{\cal A}_2$. This is the second result we advertised above: the total algebra closes exactly without central extensions. Furthermore, we note that this algebra contains normal translations, so its closure is a non-trivial statement. Since the transformations ${\cal A}_2$ considered here are the most general compatible with the presence of an isolated corner in an otherwise arbitrary $d$-dimensional bulk, $\tilde{\cal A}_2$ is the most general algebra realized by Noether charges in Einstein-Hilbert theories.\footnote{This statement concerns only diffeomorphisms  in the Einstein-Hilbert formulation of the theory. As discussed in \cite{Freidel:2020xyx, Freidel:2020svx, Freidel:2020ayo}, extensions of this algebra may arise in other formulations of gravity, due to extra symmetries on top of $\dM$.} 
In this manuscript, however, we are considering only corners at finite proper distance, because in deriving (\ref{h0}-\ref{g0}) we assumed the metric constituents to be finite when the coordinates $u^a$ go to zero. Consequently, as is clear from \eqref{g0}, the pullback of the bulk metric to the corner is uncharged under $GL(k,\RR)$, as noticed previously. One is still free to perform an arbitrary rescaling of the coordinates $u^a$, but this does not affect the metric on the corner $\gamma^{(0)}_{ij}$. 
In the case of an asymptotic corner in which the metric constituents are not strictly finite at the corner, the geometric quantities on the corner become charged under an ``extrinsic'' Weyl symmetry \cite{Ciambelli:2019bzz}, as opposed to just $Diff(S)$. 
The absence of central extensions in the charge algebra is another manifestation of this fact, which, given this discussion, could have been anticipated. We will return to a thorough study of  the interesting case of the maximal embedding algebra for asymptotic corners in future works.

Finally, let us generalize to field-dependent vector fields $\un\xi$ and $\un\eta$. In this case, the computation of the charge algebra is essentially the same, except that one should take into account  the field dependence of $\un\xi$,  as originally discussed in \cite{Barnich:2010eb}. The consequence of this is an effective modification of the Lie bracket in order for it to correctly take into account the field variations
 \beq
 [\un\eta,\un\xi]_M\equiv  [\un\eta,\un\xi]- \delta_{\un\eta}\un\xi+\delta_{\un\xi} \un\eta.
 \eeq
 Taking this effect into account through the computation of the algebra, the brackets are given by the modified Lie brackets
 \beq
\delta_{\un\eta}H_{\un\xi}=H_{[\un\xi,\un\eta]_M}=\brac{ H_{\un\xi}}{H_{\un\eta}}.
\eeq
This construction is able to describe gauge-fixed situations and/or instances with particular boundary conditions imposed. 

\subsection{Example: Near-horizon Symmetries}\label{sec:DGGP}

As an application of our general results, we show how the $BMS$-like symmetries found in \cite{Donnay:2015abr} (see also \cite{Donnay:2016ejv}) in the black hole near-horizon region are a specific instance of the maximal embedding algebra.\footnote{In this section we mainly compare with \cite{Donnay:2015abr}, but other useful references  on near horizon symmetries and null boundaries are, e.g., \cite{Penna:2015gza, Afshar:2016kjj, Hopfmuller:2016scf, Hawking:2016msc, Hawking:2016sgy, Hopfmuller:2018fni, Chandrasekaran:2018aop, Donnay:2019jiz, Carlip:2019dbu, Grumiller:2019fmp, Adami:2020amw, Chandrasekaran:2020wwn, Chen:2020nyh}.} Here, for the sake of simplicity, we will present the $d=3$ case, but the $4$-dimensional version is similar. Thus we consider a $3$-dimensional bulk geometry with a black hole, for which the near-horizon metric may be written
\beq\label{GNC}
ds^2=f dv^2+2k dv d\rho+2hdvd\phi+R^2d\phi^2.
\eeq 
In these coordinates, the horizon is located at $\rho=0$, $v$ is the null coordinate along it and $\phi$ is a $2\pi$-periodic angular coordinate. The corner we are interested in is the one embedded on the horizon at some  fixed value of $v$, say $v(\sigma)=0$, with $\sigma$ the $2\pi$-periodic coordinate on the corner. That is, calling $u^a=\{v,\rho\}$ and $x^i=\{\phi\}$, the corner is at $(u^a(\sigma)=0,\phi(\sigma)=\sigma)$, and it is at finite proper distance in the bulk, showing that this situation falls into our general treatment for the trivial embedding $\phi_0$.
Comparing \eqref{GNC} to  the parameterization \eqref{randers} gives
\beqn
h_{vv}=f,\qquad h_{\rho\rho}=0, \qquad h_{\rho v}=k,\qquad a^v_\phi=0,\qquad \gamma_{\phi\phi}=R^2,\qquad a^\rho_\phi=-{h\over k}.
\eeqn
The metric constituents in \eqref{GNC} are then expanded in powers of $\rho$, to define the solution space\footnote{Note that in our parameterization, the field $\lambda$ would be a function of $\phi$. In \cite{Donnay:2015abr}, this was written initially as $\lambda(v,\phi)$ but reduced to $\lambda(\phi)$ on shell. In that sense the on-shell solution agrees with our coordinates at linear order in $v,\rho$, and so can be compared directly.}
\beqn
f= -2\kappa \rho+O(\rho^2)&&
k= 1+O(\rho^2)\\
h= \theta(\phi)\rho+O(\rho^2)&&
R^2= \gamma(\phi)^2+\lambda(\phi)\rho+O(\rho^2),
\eeqn
from which we read the first terms in $(h_{ab},a^a_\phi,\gamma_{\phi\phi})$. The non-vanishing terms that contribute to the charges are
\beq
h^{(0)}_{\rho v}=1,\qquad h^{(1)}_{vv, \rho}=-2\kappa, \qquad a^{(1)}_\phi{}^\rho{}_\rho=-\theta(\phi),\qquad \gamma^{(0)}_{\phi\phi}=\gamma(\phi)^2.
\eeq
Furthermore, in \cite{Donnay:2015abr} the residual (that is, gauge and boundary conditions preserving) vector field $\un\chi=\chi^v\un\pa_v+\chi^\rho\un\pa_\rho+\chi^\phi\un\pa_\phi$ was found to be
\beqn
\chi^v&=&T(\phi)+O(\rho^3)\\
\chi^\rho&=&{\theta(\phi)\over 2\gamma(\phi)^2}T'(\phi)\rho^2+O(\rho^3)\\
\chi^\phi&=&Y(\phi)-{1\over \gamma(\phi)^2}T'(\phi)\rho+O(\rho^2),
\eeqn
where $T'(\phi)=\pa_\phi T(\phi)$. We therefore see that the arbitrary parameters generating the symmetries are $T(\phi)$, generating supertranslations, and $Y(\phi)$, generating superrotations, while all the other components in $\un\chi$ contain just field-dependent quantities needed in order to preserve the gauge and falloffs. Comparing this vector with the generator of the maximal embedding algebra, eq. \eqref{subalgebragens}, we  obtain\footnote{In making this comparison, since we do not require the preservation of a specific gauge of the bulk metric, we drop the higher order terms in $\un\chi$ that are there to satisfy the gauge conditions.
}
\beq
\xi^v_{(0)}=T(\phi), \qquad \xi^\rho_{(0)}=0, \qquad  \quad \xi_{(0)}^\phi=Y(\phi), \qquad \xi_{(1)}{}^a{}_b=0.
\eeq
Therefore, according to our general discussion, the subalgebra of the maximal embedding algebra realized is that of the group $ Diff(S)\ltimes \RR$, and the vector algebra is a faithful representation of the charge algebra, without central extension. This is indeed the result found in \cite{Donnay:2015abr}. There, the surface charges were computed in the covariant phase space formalism, but since fluxes were set to zero, they coincide with the Noether charge
\beq
Q(\un\chi)={1\over 16\pi G}\int_0^{2\pi}d\sigma\ \gamma(\sigma)\big[2\kappa T(\sigma)-\theta(\sigma)Y(\sigma)\big],
\eeq
agreeing exactly with \eqref{fullcharge} where, using the information gathered above (and taking $\varepsilon_{v\rho}=-1$), one computes 
\beq
N^v{}_v=1, \qquad N^\rho{}_\rho=-1, \qquad b_\phi=-\theta(\sigma), \qquad p_\rho=0, \qquad p_v=2\kappa.
\eeq
This shows how the near horizon symmetries found in \cite{Donnay:2015abr} are included in our general formalism.

\section{Conclusions}\label{sec:remarks}

In this work, we have shown that there exists a maximal field-independent closed subalgebra ${\cal A}_k$ of $\dM$ in the context of a $d$-dimensional manifold $M$ with an embedded $n$-dimensional corner $S$, with $d=n+k$. This result is an off-shell and metric-independent characterization of the Lie bracket on $M$ in the presence of $S$. We have furthermore set up the geometric framework of embeddings, showing how to adapt the bulk tangent bundle. This in turn introduces an Ehresmann connection.

It is important that these results were established without reference to a pseudo-Riemannian structure on $M$. We then discussed how the latter can be implemented, such that the Ehresmann connection appears as part of the metric constituents in $M$. The split of $TM$ to adapt to the embedding allowed us to show that, given our metric parameterization \`a la Randers-Papapetrou, the corner symmetry transforms the metric constituents as expected. The details of these transformations depend on how the metric behaves in the vicinity of the corner; in this paper, we have chosen corners at finite proper distance, while asymptotic corners require a separate analysis which we will return to in a later publication.  Using the natural non-coordinate basis in the bulk adapted to the split of $TM$, we then obtained all the various geometric data induced on $S$, together with their derivatives, which are candidate conjugate momenta.

While for generic ${\cal A}_k$, this could be of importance for higher form symmetries and higher codimensional corners, we focused our attention in the remainder of the paper on codimension-$2$. Codimension-$2$ embedded surfaces are the geometric objects on which gauge charges have support. We computed the Noether gauge charges for Einstein-Hilbert dynamics and showed that they form a representation of the subalgebra $\tilde{\cal A}_2$, where only $\mathfrak{sl}(2,\RR)$ inside $\mathfrak{gl}(2,\RR)$ is dynamically realized. The novel feature  of this analysis was to keep the full $\tilde{\cal A}_2$ symmetry in the game, including normal translations. This required a careful treatment of the embedding $\phi:S\to M$ which is an ingredient in the definition of the Noether charge. We concluded the discussion by comparing our general construction to near-horizon $BMS$-like symmetries in the absence of fluxes.

In the near future, we plan to report on how the construction can be adapted to asymptotic corners. 
There, the metric constituents are not finite, because such corners are at conformal infinity. This makes the metric constituents induced on $S$ charged under the Weyl symmetry contained in $GL(k,\RR)$, such that the embedding is actually a  {\it conformal embedding}. This analysis typically requires the specification of a normal vector field (say, tangent to an asymptotic hypersurface), and as such one could say that the algebra ${\cal A}_k$ can be regarded as (further) spontaneously broken.
Although we have focused on the gravitational Noether charge in that it gives a representation of the maximal embedding algebra on a corner, it is clearly of interest to explore other charges that have support on corners and study the effects of fluxes and edge modes.  

Finally, let us note that the truncation of $\dM$ to a closed subalgebra is reminiscent of higher spin gravity versus metric gravity, or of W-algebras versus Virasoro algebras in 2d conformal field theories.  
Codimension-2 subspaces are implicated in a wide variety of situations in (quantum) gravity, such as entanglement in the holographic context and in notions of bulk reconstruction, and thus are presumably of central interest to any serious notion of quantum gravity. We might even go so far as to interpret  $\dM$ as an emergent symmetry that arises (semi-)classically.
It may be also of interest to consider the maximal embedding algebra in the context of the S-matrix in asymptotically flat instances. 

\paragraph{Acknowledgements}
We would like to thank Francesco Alessio, Glenn Barnich, Laurent Freidel and Ali Seraj for enlightening discussions.
The research of LC was partially supported by a Marina Solvay 
Fellowship, by the ERC Advanced Grant ``High-Spin-Grav" and by 
FNRS-Belgium (convention FRFC PDR T.1025.14 and convention IISN 4.4503.15). 
The work of RGL was supported by the U.S. Department of Energy under contract DE-SC0015655.

\appendix
\renewcommand{\theequation}{\thesection.\arabic{equation}}
\setcounter{equation}{0}

\section{Levi-Civita Connection}\label{AppB}

In this Appendix we compute the bulk Levi-Civita connection and show how it induces the Levi-Civita connection of the surface, if the latter is embedded at finite distance. We will compute the spacetime connection using the non-coordinate basis $\{\un D_i,\un \pa_a\}$ in $TM$ adapted to the split introduced in Section \ref{sec2.1}. Therefore, some care must be taken as the connection coefficients will not coincide with the usual Christoffel symbols. By definition, we have
\beqn\label{definducedconn}
\LCnabla_{\un D_i}\un D_j &=& \Gamma_{ij}^k \un D_k+\Gamma_{ij}^b\un\pa_b\\
\LCnabla_{\un\pa_a}\un D_i&=& \Gamma_{ai}^j\un D_j+\Gamma_{ai}^b\un\pa_b
\\
\LCnabla_{\un D_i}\un\pa_a &=& \Gamma_{ia}^j\un D_j+\Gamma_{ia}^b\un\pa_b
\\
\LCnabla_{\un\pa_a}\un\pa_b
&=& 
\Gamma_{ab}^j\un D_j+\Gamma_{ab}^c\un\pa_c,
\eeqn
and we explicitly find
\beqn\label{indD}
\Gamma_{ja}^i=\Gamma_{aj}^i
&=&
\rho_{a}{}^i{}_j-\tfrac12h_{ab}f^b_{jk}\gamma^{ki}
\\
\Gamma_{ia}^b
&=&
\tau_{i}{}^b{}_a-\tfrac12h^{bd}(h_{ac}\varphi_i{}^c{}_d-h_{dc}\varphi_i{}^c{}_a)
\\
\Gamma_{ai}^b&=&
\tau_{i}{}^b{}_a-\tfrac12h^{bd}(h_{ac}\varphi_i{}^c{}_d+h_{dc}\varphi_i{}^c{}_a)
\\
\Gamma_{ab}^i
&=&
-\gamma^{ij}h_{ae}\tau_{j}{}^e{}_b
+\tfrac12\gamma^{ij}(h_{bc}\varphi_j{}^c{}_a+h_{ac}\varphi_j{}^c{}_b)
\\
\Gamma_{ab}^c
&=&
\tfrac12h^{cd}(\pa_ah_{db}+\pa_bh_{da}-\pa_dh_{ab}),
\eeqn
where we introduced
\beqn\label{f}
f^b_{ij}&=&D_ia_j^b-D_ja_i^b\\
\varphi_i{}^b{}_a&=&-\pa_aa^b_i\label{var}\\
\rho_{a}{}^i{}_j&=&\tfrac12\gamma^{ik}\pa_a\gamma_{kj}\label{rho}\\
\tau_{i}{}^a{}_b&=&\tfrac12h^{ac}D_ih_{cb}.\label{tau}
\eeqn
As expected, the pullback of \eqref{indD} reverts to the Christoffel connection of the induced metric on $S$, whereas the other components displayed contain information on the various geometrical quantities (connections and curvatures) for the maximal embedding algebra. The quantities (\ref{f}-\ref{tau}) gather together the various first derivative of the metric constituents. In the bulk, they would be therefore momenta with respect to certain flows. Inspired by the hydrodynamic formalism (see e.g. \cite{Damour}), if we think in terms of a normal flow generated by $\un\pa_a$, one then refers to \eqref{f} as the vorticity, \eqref{var} the acceleration and the traceless part of \eqref{rho} as the shear. 

We note that an example of this construction in codimension-1 was given in \cite{Ciambelli:2019bzz}. There, the boundary metric (rather, conformal class of metrics) was charged under Weyl, resulting in an induced Weyl connection rather than a Levi-Civita connection. In the present treatment, the surface $S$ is embedded at finite distance, and its induced metric is only charged under $Diff(S)$, plus the effects of normal translations, as shown in \eqref{g0}. Consequently, the bulk Levi-Civita connection induces the surface Levi-Civita connection. We plan to come back in future works to this question for infinite distance embedded surfaces.

\section{Changes of Embedding}\label{AppA}

We show here eq. \eqref{delphi} by computing explicitly the left-hand side and prove that it evaluates to \eqref{RHS}. This is a check of the passive versus active interpretation of $\delta_{\un\eta}$. Changing the embedding is an active point of view, whereas seeing the action of $\delta_{\un\eta}$ as a change of coordinates $y$ in the bulk for the same embedding is a passive interpretation, which is then straightforward to evaluate.

So here we work actively on the embedding, using $\phi_0$ as the initial one, and compute
\beq
(\delta_{\un\eta}\phi^*)(K[\un\xi,g,y])=\phi_{\un\teta}^*(K[\un\xi,g,y])-\phi_{0}^*(K[\un\xi,g,y]).
\eeq
In this expression we have introduced the new embedding $\phi_{\un\teta}$ corresponding to $y_{\un\teta}^M(\sigma)=y_0^M(\sigma)-\teta^M(\sigma)$, where at this point we think of this actively as a new embedding, with $y_{\un\teta}^M(\sigma)$ unrelated to a change of coordinates in the bulk. We then expand for $\un\teta$ infinitesimal and obtain
\beqn
\phi_{\un\teta}^*(K[\un\xi,g,y])&=&{1\over 2}k_{\un\xi}^{M_1 M_2}(y_{\un\teta}(\sigma)){1\over n!}\varepsilon_{M_1...M_d}dy^{M_3}_{\un\teta}(\sigma)\wedge ...\wedge dy^{M_d}_{\un\teta}(\sigma)\\
&=&{1\over 2}\Big(k_{\un\xi}^{M_1 M_2}(y_{0}(\sigma))-\teta^P(\sigma)\pa_P k_{\un\xi}^{M_1 M_2}(y_{0}(\sigma))\Big){1\over n!}\varepsilon_{M_1...M_d}dy^{M_3}_{\un\teta}(\sigma)\wedge ...\wedge dy^{M_d}_{\un\teta}(\sigma)\\
&=& {1\over 2}k_{\un\xi}^{M_1 M_2}(y_{0}(\sigma)){1\over n!}\varepsilon_{M_1...M_d}dy^{M_3}_{\un\teta}(\sigma)\wedge ...\wedge dy^{M_d}_{\un\teta}(\sigma)\\
&&-{1\over 2}\teta^P(\sigma)\pa_P k_{\un\xi}^{M_1 M_2}(y_{0}(\sigma)){1\over n!}\varepsilon_{M_1...M_d}dy^{M_3}_{0}(\sigma)\wedge ...\wedge dy^{M_d}_{0}(\sigma)\nonumber\\
&=&\phi_{0}^*(K[\un\xi,g,y])-{1\over 2}\teta^P(\sigma)\pa_P k_{\un\xi}^{ab}(y_{0}(\sigma))\varepsilon_{ab}{1\over n!}\varepsilon_{i_1..i_n}(\delta_{\alpha_1}^{i_1}...\delta_{\alpha_n}^{i_n})d\sigma^{\alpha_1}\wedge ...\wedge d\sigma^{\alpha_n}\\
&&- {1\over 2}k_{\un\xi}^{M_1 M_2}(y_{0}(\sigma)){1\over (n-1)!}\varepsilon_{M_1M_2M_3i_2...i_n} \pa_{\alpha_1}\teta^{M_3}(\sigma)(\delta_{\alpha_2}^{i_2}...\delta_{\alpha_n}^{i_n})d\sigma^{\alpha_1}\wedge d\sigma^{\alpha_2}\wedge ...\wedge d\sigma^{\alpha_n}\nonumber\\
&=&\phi_{0}^*(K[\un\xi,g,y])-{1\over 2}\teta^P(\sigma)\pa_P k_{\un\xi}^{ab}(y_{0}(\sigma))\varepsilon_{ab}vol_0\\
&&- {1\over 2}k_{\un\xi}^{ab}(y_{0}(\sigma))\varepsilon_{ab}{1\over (n-1)!}\varepsilon_{ii_2...i_n} \pa_{\alpha_1}\teta^{i}(\sigma)(\delta_{\alpha_2}^{i_2}...\delta_{\alpha_n}^{i_n})d\sigma^{\alpha_1}\wedge d\sigma^{\alpha_2}\wedge ...\wedge d\sigma^{\alpha_n}\nonumber\\
&&- k_{\un\xi}^{ja}(y_{0}(\sigma)){1\over (n-1)!}\varepsilon_{ab}\varepsilon_{ii_2...i_n} \pa_{\alpha_1}\teta^{b}(\sigma)(\delta_{\alpha_2}^{i_2}...\delta_{\alpha_n}^{i_n})d\sigma^{\alpha_1}\wedge d\sigma^{\alpha_2}\wedge ...\wedge d\sigma^{\alpha_n}\nonumber.
\eeqn
Subtracting $\phi_{0}^*(K[\un\xi,g,y])$ and using the identity ${1\over (n-1)!}\varepsilon_{ii_2...i_n}(\delta_{\alpha_2}^{i_2}...\delta_{\alpha_n}^{i_n})d\sigma^{\alpha_1}\wedge d\sigma^{\alpha_2}\wedge ...\wedge d\sigma^{\alpha_n}=\delta^{\alpha_1}_i vol_0$ we find
\beqn
(\delta_{\un\eta}\phi^*)(K[\un\xi,g,y])=-\Big[{1\over 2}\teta^P(\sigma)\pa_P k_{\un\xi}^{ab}(y_{0}(\sigma))\varepsilon_{ab}+ {1\over 2}k_{\un\xi}^{ab}(y_{0}(\sigma))\varepsilon_{ab}\pa_{\alpha_1}\teta^{i}(\sigma)\delta^{\alpha_1}_{i}+k_{\un\xi}^{ja}(y_{0}(\sigma))\varepsilon_{ab}\pa_{\alpha_1}\teta^{b}(\sigma)\delta^{\alpha_1}_{j}\Big]vol_0\nonumber.
\eeqn
Note that the last term has an interpretation in terms of the pullback flat connection $a_\alpha^b=-\pa_\alpha\teta^b(\sigma)$ appropriate to the embedding $\phi_{\teta}$.
To then make contact with the passive interpretation, we relate the change of embedding to a change of coordinates in the bulk. This is achieved if we identify $\teta^M(\sigma)=\eta^M(y_0(\sigma))$, such that, if the new coordinates in the bulk are $y'^M=y^M-\eta^M(y)$, then we simply have $y'^M(\sigma)=y_{\un\teta}^M(\sigma)$. From this follows $\pa_{\alpha_1} \teta^{M}(\sigma)=\pa_{\alpha_1} \eta^{M}(y_0(\sigma))=\pa_{\alpha_1}u^a(\sigma)\pa_a\eta^M(y_0(\sigma))+\pa_{\alpha_1}x^i(\sigma)\pa_i\eta^M(y_0(\sigma))\simeq\delta_{\alpha_1}^i\pa_i\eta^M(y_0(\sigma))$ and so we obtain
\beqn
(\delta_{\un\eta}\phi^*)(K[\un\xi,g,y])&=&-\Big[{1\over 2}\eta^P(y_0(\sigma))\pa_P k_{\un\xi}^{ab}(y_{0}(\sigma))\varepsilon_{ab}+ {1\over 2}k_{\un\xi}^{ab}(y_{0}(\sigma))\varepsilon_{ab}\pa_{i}\eta^{i}(y_0(\sigma))+k_{\un\xi}^{ja}(y_{0}(\sigma))\varepsilon_{ab}\pa_{j}\eta^{b}(y_0(\sigma))\Big]vol_0\nonumber\\
&=&-\Big[
\varepsilon_{ab}k_{\un\xi}^{ja}\pa_{j}\eta^b 
+\frac12 \eta^c\pa_c(\varepsilon_{ab}k_{\un\xi}^{ab})
+\frac12 \pa_{j}\Big(\eta^j\varepsilon_{ab}k_{\un\xi}^{ab}\Big)
\Big](y_{0}(\sigma))\ vol_0.
\eeqn
This result is exactly \eqref{RHS}, proving thus eq. \eqref{delphi}, which is a crucial step in the derivation of the Noether charge algebra.


\begin{thebibliography}{10}

\bibitem{Arnowitt:1962hi}
R.~L. Arnowitt, S.~Deser, and C.~W. Misner, ``{The Dynamics of general
  relativity},'' \href{http://dx.doi.org/10.1007/s10714-008-0661-1}{{\em Gen.
  Rel. Grav.} {\bf 40} (2008)  1997--2027},
  \href{http://arxiv.org/abs/gr-qc/0405109}{{\tt arXiv:gr-qc/0405109}}.

\bibitem{Regge:1974zd}
T.~Regge and C.~Teitelboim, ``{Role of Surface Integrals in the Hamiltonian
  Formulation of General Relativity},''
  \href{http://dx.doi.org/10.1016/0003-4916(74)90404-7}{{\em Annals Phys.} {\bf
  88} (1974)  286}.

\bibitem{Henneaux:1985tv}
M.~Henneaux and C.~Teitelboim, ``{Asymptotically anti-De Sitter Spaces},''
  \href{http://dx.doi.org/10.1007/BF01205790}{{\em Commun. Math. Phys.} {\bf
  98} (1985)  391--424}.

\bibitem{Brown:1986ed}
J.~D. Brown and M.~Henneaux, ``{On the Poisson Brackets of Differentiable
  Generators in Classical Field Theory},''
  \href{http://dx.doi.org/10.1063/1.527249}{{\em J. Math. Phys.} {\bf 27}
  (1986)  489--491}.

\bibitem{Brown:1986nw}
J.~Brown and M.~Henneaux, ``{Central Charges in the Canonical Realization of
  Asymptotic Symmetries: An Example from Three-Dimensional Gravity},''
  \href{http://dx.doi.org/10.1007/BF01211590}{{\em Commun.\ Math.\ Phys.} {\bf
  104} (1986)  207--226}.

\bibitem{Kijowski:1976ze}
J.~Kijowski and W.~Szczyrba, ``{A Canonical Structure for Classical Field
  Theories},'' \href{http://dx.doi.org/10.1007/BF01608496}{{\em Commun. Math.
  Phys.} {\bf 46} (1976)  183--206}.

\bibitem{Crnkovic:1986ex}
C.~Crnkovic and E.~Witten, {\em Three Hundred Years of Gravitation},
  ch.~{Covariant Description of Canonical Formalism in Geometrical Theories},
  pp.~676--684.
\newblock Cambridge University Press, 1986.

\bibitem{Lee:1990nz}
J.~Lee and R.~M. Wald, ``{Local symmetries and constraints},''
  \href{http://dx.doi.org/10.1063/1.528801}{{\em J. Math. Phys.} {\bf 31}
  (1990)  725--743}.

\bibitem{ASHTEKAR1991417}
A.~Ashtekar, L.~Bombelli, and O.~Reula,
  \href{http://dx.doi.org/https://doi.org/10.1016/B978-0-444-88958-4.50021-5}{``The
  covariant phase space of asymptotically flat gravitational fields,''} in {\em
  Mechanics, Analysis and Geometry: 200 Years After Lagrange}, M.~Francaviglia,
  ed., North-Holland Delta Series, pp.~417--450.
\newblock Elsevier, Amsterdam, 1991.

\bibitem{Wald:1993nt}
R.~M. Wald, ``{Black hole entropy is the Noether charge},''
  \href{http://dx.doi.org/10.1103/PhysRevD.48.R3427}{{\em Phys. Rev. D} {\bf
  48} (1993) no.~8, 3427--3431}, \href{http://arxiv.org/abs/gr-qc/9307038}{{\tt
  arXiv:gr-qc/9307038}}.

\bibitem{Iyer:1994ys}
V.~Iyer and R.~M. Wald, ``{Some properties of Noether charge and a proposal for
  dynamical black hole entropy},''
  \href{http://dx.doi.org/10.1103/PhysRevD.50.846}{{\em Phys. Rev. D} {\bf 50}
  (1994)  846--864}, \href{http://arxiv.org/abs/gr-qc/9403028}{{\tt
  arXiv:gr-qc/9403028}}.

\bibitem{Wald:1999wa}
R.~M. Wald and A.~Zoupas, ``{A General definition of 'conserved quantities' in
  general relativity and other theories of gravity},''
  \href{http://dx.doi.org/10.1103/PhysRevD.61.084027}{{\em Phys. Rev. D} {\bf
  61} (2000)  084027}, \href{http://arxiv.org/abs/gr-qc/9911095}{{\tt
  arXiv:gr-qc/9911095}}.

\bibitem{doi:10.1080/00411457108231446}
E.~Noether, ``Invariant variation problems,''
  \href{http://dx.doi.org/10.1080/00411457108231446}{{\em Transport Theory and
  Statistical Physics} {\bf 1} (1971) no.~3, 186--207}.

\bibitem{Barnich:1995ap}
G.~Barnich, F.~Brandt, and M.~Henneaux, ``{Local BRST cohomology in
  Einstein-Yang-Mills theory},''
  \href{http://dx.doi.org/10.1016/0550-3213(95)00471-4}{{\em Nucl. Phys. B}
  {\bf 455} (1995)  357--408}, \href{http://arxiv.org/abs/hep-th/9505173}{{\tt
  arXiv:hep-th/9505173}}.

\bibitem{Barnich:2000zw}
G.~Barnich, F.~Brandt, and M.~Henneaux, ``{Local BRST cohomology in gauge
  theories},'' \href{http://dx.doi.org/10.1016/S0370-1573(00)00049-1}{{\em
  Phys. Rept.} {\bf 338} (2000)  439--569},
  \href{http://arxiv.org/abs/hep-th/0002245}{{\tt arXiv:hep-th/0002245}}.

\bibitem{Barnich:2001jy}
G.~Barnich and F.~Brandt, ``{Covariant theory of asymptotic symmetries,
  conservation laws and central charges},''
  \href{http://dx.doi.org/10.1016/S0550-3213(02)00251-1}{{\em Nucl. Phys. B}
  {\bf 633} (2002)  3--82}, \href{http://arxiv.org/abs/hep-th/0111246}{{\tt
  arXiv:hep-th/0111246}}.

\bibitem{Barnich:2007bf}
G.~Barnich and G.~Compere, ``{Surface charge algebra in gauge theories and
  thermodynamic integrability},''
  \href{http://dx.doi.org/10.1063/1.2889721}{{\em J. Math. Phys.} {\bf 49}
  (2008)  042901}, \href{http://arxiv.org/abs/0708.2378}{{\tt arXiv:0708.2378
  [gr-qc]}}.

\bibitem{Freidel:2015gpa}
L.~Freidel and A.~Perez, ``{Quantum gravity at the corner},''
  \href{http://dx.doi.org/10.3390/universe4100107}{{\em Universe} {\bf 4}
  (2018) no.~10, 107}, \href{http://arxiv.org/abs/1507.02573}{{\tt
  arXiv:1507.02573 [gr-qc]}}.

\bibitem{Donnelly:2016auv}
W.~Donnelly and L.~Freidel, ``{Local subsystems in gauge theory and gravity},''
  \href{http://dx.doi.org/10.1007/JHEP09(2016)102}{{\em JHEP} {\bf 09} (2016)
  102}, \href{http://arxiv.org/abs/1601.04744}{{\tt arXiv:1601.04744
  [hep-th]}}.

\bibitem{Freidel:2020xyx}
L.~Freidel, M.~Geiller, and D.~Pranzetti, ``{Edge modes of gravity. Part I.
  Corner potentials and charges},''
  \href{http://dx.doi.org/10.1007/JHEP11(2020)026}{{\em JHEP} {\bf 11} (2020)
  026}, \href{http://arxiv.org/abs/2006.12527}{{\tt arXiv:2006.12527
  [hep-th]}}.

\bibitem{Freidel:2020svx}
L.~Freidel, M.~Geiller, and D.~Pranzetti, ``{Edge modes of gravity. Part II.
  Corner metric and Lorentz charges},''
  \href{http://dx.doi.org/10.1007/JHEP11(2020)027}{{\em JHEP} {\bf 11} (2020)
  027}, \href{http://arxiv.org/abs/2007.03563}{{\tt arXiv:2007.03563
  [hep-th]}}.

\bibitem{Freidel:2020ayo}
L.~Freidel, M.~Geiller, and D.~Pranzetti, ``{Edge modes of gravity. Part III.
  Corner simplicity constraints},''
  \href{http://dx.doi.org/10.1007/JHEP01(2021)100}{{\em JHEP} {\bf 01} (2021)
  100}, \href{http://arxiv.org/abs/2007.12635}{{\tt arXiv:2007.12635
  [hep-th]}}.

\bibitem{Donnelly:2014fua}
W.~Donnelly and A.~C. Wall, ``{Entanglement entropy of electromagnetic edge
  modes},'' \href{http://dx.doi.org/10.1103/PhysRevLett.114.111603}{{\em Phys.
  Rev. Lett.} {\bf 114} (2015) no.~11, 111603},
  \href{http://arxiv.org/abs/1412.1895}{{\tt arXiv:1412.1895 [hep-th]}}.

\bibitem{Speranza:2017gxd}
A.~J. Speranza, ``{Local Phase Space and Edge Modes for
  Diffeomorphism-invariant Theories},''
  \href{http://dx.doi.org/10.1007/JHEP02(2018)021}{{\em JHEP} {\bf 02} (2018)
  021}, \href{http://arxiv.org/abs/1706.05061}{{\tt arXiv:1706.05061
  [hep-th]}}.

\bibitem{Donnelly:2020xgu}
W.~Donnelly, L.~Freidel, S.~F. Moosavian, and A.~J. Speranza, ``{Gravitational
  Edge Modes, Coadjoint Orbits, and Hydrodynamics},''
  \href{http://arxiv.org/abs/2012.10367}{{\tt arXiv:2012.10367 [hep-th]}}.

\bibitem{Ciambelli:2019bzz}
L.~Ciambelli and R.~G. Leigh, ``{Weyl Connections and their Role in
  Holography},'' \href{http://dx.doi.org/10.1103/PhysRevD.101.086020}{{\em
  Phys. Rev. D} {\bf 101} (2020) no.~8, 086020},
  \href{http://arxiv.org/abs/1905.04339}{{\tt arXiv:1905.04339 [hep-th]}}.

\bibitem{Ciambelli:2019lap}
L.~Ciambelli, R.~G. Leigh, C.~Marteau, and P.~M. Petropoulos, ``{Carroll
  Structures, Null Geometry and Conformal Isometries},''
  \href{http://dx.doi.org/10.1103/PhysRevD.100.046010}{{\em Phys. Rev. D} {\bf
  100} (2019) no.~4, 046010}, \href{http://arxiv.org/abs/1905.02221}{{\tt
  arXiv:1905.02221 [hep-th]}}.

\bibitem{PhysRev.59.195}
G.~Randers, \href{http://dx.doi.org/10.1103/PhysRev.59.195}{``On an
  asymmetrical metric in the four-space of general relativity,''{\em Phys.
  Rev.} {\bf 59} (Jan, 1941)  195--199}.
  \url{https://link.aps.org/doi/10.1103/PhysRev.59.195}.

\bibitem{Papapetrou:1966zz}
A.~Papapetrou, ``{Champs gravitationnels stationnaires a symetrie axiale},''
  {\em Ann. Inst. H. Poincare Phys. Theor.} {\bf 4} (1966)  83--105.

\bibitem{Gibbons:2008zi}
G.~W. Gibbons, C.~A.~R. Herdeiro, C.~M. Warnick, and M.~C. Werner,
  ``{Stationary Metrics and Optical Zermelo-Randers-Finsler Geometry},''
  \href{http://dx.doi.org/10.1103/PhysRevD.79.044022}{{\em Phys. Rev. D} {\bf
  79} (2009)  044022}, \href{http://arxiv.org/abs/0811.2877}{{\tt
  arXiv:0811.2877 [gr-qc]}}.

\bibitem{https://doi.org/10.1002/zamm.19310110205}
E.~Zermelo, ``{\"U}ber das navigationsproblem bei ruhender oder
  ver{\"a}nderlicher windverteilung,''
  \href{http://dx.doi.org/https://doi.org/10.1002/zamm.19310110205}{{\em ZAMM -
  Journal of Applied Mathematics and Mechanics / Zeitschrift f{\"u}r Angewandte
  Mathematik und Mechanik} {\bf 11} (1931) no.~2, 114--124}.

\bibitem{Speranza:2019hkr}
A.~J. Speranza, ``{Geometrical tools for embedding fields, submanifolds, and
  foliations},'' \href{http://arxiv.org/abs/1904.08012}{{\tt arXiv:1904.08012
  [gr-qc]}}.

\bibitem{Kapec:2014opa}
D.~Kapec, V.~Lysov, S.~Pasterski, and A.~Strominger, ``{Semiclassical Virasoro
  symmetry of the quantum gravity $ \mathcal{S}$-matrix},''
  \href{http://dx.doi.org/10.1007/JHEP08(2014)058}{{\em JHEP} {\bf 08} (2014)
  058}, \href{http://arxiv.org/abs/1406.3312}{{\tt arXiv:1406.3312 [hep-th]}}.

\bibitem{Campiglia:2015yka}
M.~Campiglia and A.~Laddha, ``{New symmetries for the Gravitational
  S-matrix},'' \href{http://dx.doi.org/10.1007/JHEP04(2015)076}{{\em JHEP} {\bf
  04} (2015)  076}, \href{http://arxiv.org/abs/1502.02318}{{\tt
  arXiv:1502.02318 [hep-th]}}.

\bibitem{Pasterski:2016qvg}
S.~Pasterski, S.-H. Shao, and A.~Strominger, ``{Flat Space Amplitudes and
  Conformal Symmetry of the Celestial Sphere},''
  \href{http://dx.doi.org/10.1103/PhysRevD.96.065026}{{\em Phys. Rev. D} {\bf
  96} (2017) no.~6, 065026}, \href{http://arxiv.org/abs/1701.00049}{{\tt
  arXiv:1701.00049 [hep-th]}}.

\bibitem{Strominger:2017zoo}
A.~Strominger, ``{Lectures on the Infrared Structure of Gravity and Gauge
  Theory},'' \href{http://arxiv.org/abs/1703.05448}{{\tt arXiv:1703.05448
  [hep-th]}}.

\bibitem{Donnay:2020guq}
L.~Donnay, S.~Pasterski, and A.~Puhm, ``{Asymptotic Symmetries and Celestial
  CFT},'' \href{http://dx.doi.org/10.1007/JHEP09(2020)176}{{\em JHEP} {\bf 09}
  (2020)  176}, \href{http://arxiv.org/abs/2005.08990}{{\tt arXiv:2005.08990
  [hep-th]}}.

\bibitem{Penrose:1962ij}
R.~Penrose, ``{Asymptotic properties of fields and space-times},''
  \href{http://dx.doi.org/10.1103/PhysRevLett.10.66}{{\em Phys. Rev. Lett.}
  {\bf 10} (1963)  66--68}.

\bibitem{Carter:1966zza}
B.~Carter, ``{Complete Analytic Extension of the Symmetry Axis of Kerr's
  Solution of Einstein's Equations},''
  \href{http://dx.doi.org/10.1103/PhysRev.141.1242}{{\em Phys. Rev.} {\bf 141}
  (1966)  1242--1247}.

\bibitem{Hawking:1973uf}
S.~W. Hawking and G.~F.~R. Ellis,
  \href{http://dx.doi.org/10.1017/CBO9780511524646}{{\em {The Large Scale
  Structure of Space-Time}}}.
\newblock Cambridge Monographs on Mathematical Physics. Cambridge University
  Press, 2, 2011.

\bibitem{Newman:1961qr}
E.~Newman and R.~Penrose, ``{An Approach to gravitational radiation by a method
  of spin coefficients},'' \href{http://dx.doi.org/10.1063/1.1724257}{{\em J.
  Math. Phys.} {\bf 3} (1962)  566--578}.

\bibitem{Geroch:1973am}
R.~P. Geroch, A.~Held, and R.~Penrose, ``{A space-time calculus based on pairs
  of null directions},'' \href{http://dx.doi.org/10.1063/1.1666410}{{\em J.
  Math. Phys.} {\bf 14} (1973)  874--881}.

\bibitem{Donnay:2015abr}
L.~Donnay, G.~Giribet, H.~A. Gonzalez, and M.~Pino, ``{Supertranslations and
  Superrotations at the Black Hole Horizon},''
  \href{http://dx.doi.org/10.1103/PhysRevLett.116.091101}{{\em Phys. Rev.
  Lett.} {\bf 116} (2016) no.~9, 091101},
  \href{http://arxiv.org/abs/1511.08687}{{\tt arXiv:1511.08687 [hep-th]}}.

\bibitem{Campiglia:2014yka}
M.~Campiglia and A.~Laddha, ``{Asymptotic symmetries and subleading soft
  graviton theorem},'' \href{http://dx.doi.org/10.1103/PhysRevD.90.124028}{{\em
  Phys. Rev. D} {\bf 90} (2014) no.~12, 124028},
  \href{http://arxiv.org/abs/1408.2228}{{\tt arXiv:1408.2228 [hep-th]}}.

\bibitem{Compere:2018ylh}
G.~Comp\`ere, A.~Fiorucci, and R.~Ruzziconi, ``{Superboost transitions,
  refraction memory and super-Lorentz charge algebra},''
  \href{http://dx.doi.org/10.1007/JHEP11(2018)200}{{\em JHEP} {\bf 11} (2018)
  200}, \href{http://arxiv.org/abs/1810.00377}{{\tt arXiv:1810.00377
  [hep-th]}}. [Erratum: JHEP 04, 172 (2020)].

\bibitem{doi:10.1098/rspa.1962.0161}
H.~Bondi, M.~G.~J. Van~der Burg, and A.~W.~K. Metzner, ``{Gravitational waves
  in general relativity, VII. Waves from axi-symmetric isolated system},''
  \href{http://dx.doi.org/10.1098/rspa.1962.0161}{{\em Proceedings of the Royal
  Society of London. Series A. Mathematical and Physical Sciences} {\bf 269}
  (1962) no.~1336, 21--52}.

\bibitem{doi:10.1098/rspa.1962.0206}
R.~Sachs and H.~Bondi, ``{Gravitational Waves in General Relativity VIII. Waves
  in Asymptotically Flat Space-time},''
  \href{http://dx.doi.org/10.1098/rspa.1962.0206}{{\em Proceedings of the Royal
  Society of London. Series A. Mathematical and Physical Sciences} {\bf 270}
  (1962) no.~1340, 103--126}.

\bibitem{Barnich:2010eb}
G.~Barnich and C.~Troessaert, ``{Aspects of the BMS/CFT correspondence},''
  \href{http://dx.doi.org/10.1007/JHEP05(2010)062}{{\em JHEP} {\bf 05} (2010)
  062}, \href{http://arxiv.org/abs/1001.1541}{{\tt arXiv:1001.1541 [hep-th]}}.

\bibitem{Compere:2019bua}
G.~Comp\`ere, A.~Fiorucci, and R.~Ruzziconi, ``{The $\Lambda$-BMS$_4$ group of
  dS$_4$ and new boundary conditions for AdS$_4$},''
  \href{http://dx.doi.org/10.1088/1361-6382/ab3d4b}{{\em Class. Quant. Grav.}
  {\bf 36} (2019) no.~19, 195017}, \href{http://arxiv.org/abs/1905.00971}{{\tt
  arXiv:1905.00971 [gr-qc]}}.

\bibitem{Freidel:2021yqe}
L.~Freidel, R.~Oliveri, D.~Pranzetti, and S.~Speziale, ``{The Weyl BMS group
  and Einstein's equations},'' \href{http://arxiv.org/abs/2104.05793}{{\tt
  arXiv:2104.05793 [hep-th]}}.

\bibitem{Adami:2020ugu}
H.~Adami, M.~M. Sheikh-Jabbari, V.~Taghiloo, H.~Yavartanoo, and C.~Zwikel,
  ``{Symmetries at null boundaries: two and three dimensional gravity cases},''
  \href{http://dx.doi.org/10.1007/JHEP10(2020)107}{{\em JHEP} {\bf 10} (2020)
  107}, \href{http://arxiv.org/abs/2007.12759}{{\tt arXiv:2007.12759
  [hep-th]}}.

\bibitem{Adamo:2009vu}
T.~M. Adamo, C.~N. Kozameh, and E.~T. Newman, ``{Null Geodesic Congruences,
  Asymptotically Flat Space-Times and Their Physical Interpretation},''
  \href{http://dx.doi.org/10.12942/lrr-2009-6}{{\em Living Rev. Rel.} {\bf 12}
  (2009)  6}, \href{http://arxiv.org/abs/0906.2155}{{\tt arXiv:0906.2155
  [gr-qc]}}.

\bibitem{Penrose:2011wy}
R.~Penrose, ``Republication of: Conformal treatment of infinity,''
  \href{http://dx.doi.org/10.1007/s10714-010-1110-5}{{\em General Relativity
  and Gravitation} {\bf 43} (2011) no.~3, 901--922}.

\bibitem{Alessio:2020ioh}
F.~Alessio, G.~Barnich, L.~Ciambelli, P.~Mao, and R.~Ruzziconi, ``{Weyl charges
  in asymptotically locally AdS$_3$ spacetimes},''
  \href{http://dx.doi.org/10.1103/PhysRevD.103.046003}{{\em Phys. Rev. D} {\bf
  103} (2021) no.~4, 046003}, \href{http://arxiv.org/abs/2010.15452}{{\tt
  arXiv:2010.15452 [hep-th]}}.

\bibitem{Freidel:2021cbc}
L.~Freidel, R.~Oliveri, D.~Pranzetti, and S.~Speziale, ``{Extended corner
  symmetry, charge bracket and Einstein's equations},''
  \href{http://arxiv.org/abs/2104.12881}{{\tt arXiv:2104.12881 [hep-th]}}.

\bibitem{Barnich:2015uva}
G.~Barnich and B.~Oblak, ``{Notes on the BMS group in three dimensions: II.
  Coadjoint representation},''
  \href{http://dx.doi.org/10.1007/JHEP03(2015)033}{{\em JHEP} {\bf 03} (2015)
  033}, \href{http://arxiv.org/abs/1502.00010}{{\tt arXiv:1502.00010
  [hep-th]}}.

\bibitem{Barnich:2021dta}
G.~Barnich and R.~Ruzziconi, ``{Coadjoint representation of the BMS group on
  celestial Riemann surfaces},'' \href{http://arxiv.org/abs/2103.11253}{{\tt
  arXiv:2103.11253 [gr-qc]}}.

\bibitem{Freidel:2019ofr}
L.~Freidel, E.~R. Livine, and D.~Pranzetti, ``{Kinematical Gravitational Charge
  Algebra},'' \href{http://dx.doi.org/10.1103/PhysRevD.101.024012}{{\em Phys.
  Rev. D} {\bf 101} (2020) no.~2, 024012},
  \href{http://arxiv.org/abs/1910.05642}{{\tt arXiv:1910.05642 [gr-qc]}}.

\bibitem{Donnay:2016ejv}
L.~Donnay, G.~Giribet, H.~A. Gonz\'alez, and M.~Pino, ``{Extended Symmetries at
  the Black Hole Horizon},''
  \href{http://dx.doi.org/10.1007/JHEP09(2016)100}{{\em JHEP} {\bf 09} (2016)
  100}, \href{http://arxiv.org/abs/1607.05703}{{\tt arXiv:1607.05703
  [hep-th]}}.

\bibitem{Penna:2015gza}
R.~F. Penna, ``{BMS invariance and the membrane paradigm},''
  \href{http://dx.doi.org/10.1007/JHEP03(2016)023}{{\em JHEP} {\bf 03} (2016)
  023}, \href{http://arxiv.org/abs/1508.06577}{{\tt arXiv:1508.06577
  [hep-th]}}.

\bibitem{Afshar:2016kjj}
H.~Afshar, D.~Grumiller, W.~Merbis, A.~Perez, D.~Tempo, and R.~Troncoso,
  ``{Soft hairy horizons in three spacetime dimensions},''
  \href{http://dx.doi.org/10.1103/PhysRevD.95.106005}{{\em Phys. Rev. D} {\bf
  95} (2017) no.~10, 106005}, \href{http://arxiv.org/abs/1611.09783}{{\tt
  arXiv:1611.09783 [hep-th]}}.

\bibitem{Hopfmuller:2016scf}
F.~Hopfm\"uller and L.~Freidel, ``{Gravity Degrees of Freedom on a Null
  Surface},'' \href{http://dx.doi.org/10.1103/PhysRevD.95.104006}{{\em Phys.
  Rev. D} {\bf 95} (2017) no.~10, 104006},
  \href{http://arxiv.org/abs/1611.03096}{{\tt arXiv:1611.03096 [gr-qc]}}.

\bibitem{Hawking:2016msc}
S.~W. Hawking, M.~J. Perry, and A.~Strominger, ``{Soft Hair on Black Holes},''
  \href{http://dx.doi.org/10.1103/PhysRevLett.116.231301}{{\em Phys. Rev.
  Lett.} {\bf 116} (2016) no.~23, 231301},
  \href{http://arxiv.org/abs/1601.00921}{{\tt arXiv:1601.00921 [hep-th]}}.

\bibitem{Hawking:2016sgy}
S.~W. Hawking, M.~J. Perry, and A.~Strominger, ``{Superrotation Charge and
  Supertranslation Hair on Black Holes},''
  \href{http://dx.doi.org/10.1007/JHEP05(2017)161}{{\em JHEP} {\bf 05} (2017)
  161}, \href{http://arxiv.org/abs/1611.09175}{{\tt arXiv:1611.09175
  [hep-th]}}.

\bibitem{Hopfmuller:2018fni}
F.~Hopfm\"uller and L.~Freidel, ``{Null Conservation Laws for Gravity},''
  \href{http://dx.doi.org/10.1103/PhysRevD.97.124029}{{\em Phys. Rev. D} {\bf
  97} (2018) no.~12, 124029}, \href{http://arxiv.org/abs/1802.06135}{{\tt
  arXiv:1802.06135 [gr-qc]}}.

\bibitem{Chandrasekaran:2018aop}
V.~Chandrasekaran, E.~E. Flanagan, and K.~Prabhu, ``{Symmetries and charges of
  general relativity at null boundaries},''
  \href{http://dx.doi.org/10.1007/JHEP11(2018)125}{{\em JHEP} {\bf 11} (2018)
  125}, \href{http://arxiv.org/abs/1807.11499}{{\tt arXiv:1807.11499
  [hep-th]}}.

\bibitem{Donnay:2019jiz}
L.~Donnay and C.~Marteau, ``{Carrollian Physics at the Black Hole Horizon},''
  \href{http://dx.doi.org/10.1088/1361-6382/ab2fd5}{{\em Class. Quant. Grav.}
  {\bf 36} (2019) no.~16, 165002}, \href{http://arxiv.org/abs/1903.09654}{{\tt
  arXiv:1903.09654 [hep-th]}}.

\bibitem{Carlip:2019dbu}
S.~Carlip, ``{Near-horizon Bondi-Metzner-Sachs symmetry, dimensional reduction,
  and black hole entropy},''
  \href{http://dx.doi.org/10.1103/PhysRevD.101.046002}{{\em Phys. Rev. D} {\bf
  101} (2020) no.~4, 046002}, \href{http://arxiv.org/abs/1910.01762}{{\tt
  arXiv:1910.01762 [hep-th]}}.

\bibitem{Grumiller:2019fmp}
D.~Grumiller, A.~P\'erez, M.~M. Sheikh-Jabbari, R.~Troncoso, and C.~Zwikel,
  ``{Spacetime structure near generic horizons and soft hair},''
  \href{http://dx.doi.org/10.1103/PhysRevLett.124.041601}{{\em Phys. Rev.
  Lett.} {\bf 124} (2020) no.~4, 041601},
  \href{http://arxiv.org/abs/1908.09833}{{\tt arXiv:1908.09833 [hep-th]}}.

\bibitem{Adami:2020amw}
H.~Adami, D.~Grumiller, S.~Sadeghian, M.~M. Sheikh-Jabbari, and C.~Zwikel,
  ``{T-Witts from the horizon},''
  \href{http://dx.doi.org/10.1007/JHEP04(2020)128}{{\em JHEP} {\bf 04} (2020)
  128}, \href{http://arxiv.org/abs/2002.08346}{{\tt arXiv:2002.08346
  [hep-th]}}.

\bibitem{Chandrasekaran:2020wwn}
V.~Chandrasekaran and A.~J. Speranza, ``{Anomalies in gravitational charge
  algebras of null boundaries and black hole entropy},''
  \href{http://dx.doi.org/10.1007/JHEP01(2021)137}{{\em JHEP} {\bf 01} (2021)
  137}, \href{http://arxiv.org/abs/2009.10739}{{\tt arXiv:2009.10739
  [hep-th]}}.

\bibitem{Chen:2020nyh}
L.-Q. Chen, W.~Z. Chua, S.~Liu, A.~J. Speranza, and B.~d. S.~L. Torres,
  ``{Virasoro hair and entropy for axisymmetric Killing horizons},''
  \href{http://dx.doi.org/10.1103/PhysRevLett.125.241302}{{\em Phys. Rev.
  Lett.} {\bf 125} (2020)  241302}, \href{http://arxiv.org/abs/2006.02430}{{\tt
  arXiv:2006.02430 [hep-th]}}.

\bibitem{Damour}
T.~Damour, ``{Quelques propri\'et\'es m\'ecaniques, \'electromagn\'etiques,
  thermodynamiques et quantiques des trous noirs},'' {\em Th\'ese de Doctorat
  d'Etat, Universit\'e Pierre et Marie Curie, Paris VI} (1979)  .

\end{thebibliography}

\providecommand{\href}[2]{#2}\begingroup\raggedright\endgroup

\end{document}